\documentclass[a4paper,11pt]{article}
\pdfoutput=1 % if your are submitting a pdflatex (i.e. if you have
\usepackage{jcappub} % for details on the use of the package, please
\usepackage{xcolor}

\usepackage{graphicx}% Include figure files
\usepackage{dcolumn}% Align table columns on decimal point
\usepackage{bm}% bold math
\usepackage{latexsym} 
\usepackage{amsfonts}
\usepackage{amssymb}
\usepackage{amsmath}
\usepackage{bbold}
\usepackage{datetime}
\usepackage{url}
                    
\usepackage[utf8]{inputenc}

\def\cs2{c_{s}^{2}}

\definecolor{verde}{rgb}{0,0.5,0}
\definecolor{bordeaux}{rgb}{0.5, 0, 0.12}

 \def\be   {\begin{equation}}   \def\ee   {\end{equation}}
 \def\ba   {\begin{array}}      \def\ea   {\end{array}}
 \def\bea  {\begin{eqnarray}}   \def\eea  {\end{eqnarray}}
 \def\bean {\begin{eqnarray*}}  \def\eean {\end{eqnarray*}}
\title{\boldmath Spinning Guest Fields during Inflation: Leftover Signatures}

\author[a,b]{Emanuela Dimastrogiovanni,}
\author[c,d]{Matteo Fasiello}
\author[d]{and A. Emir G\"umr\"uk\c{c}\"uo\u{g}lu}

\affiliation[a]{Van Swinderen Institute for Particle Physics and Gravity, University of Groningen, Nijenborgh 4, 9747 AG Groningen, The Netherlands}
\affiliation[b]{Sydney Consortium for Particle Physics and Cosmology, School of Physics, The  University of New South Wales, Sydney NSW 2052, Australia}
\affiliation[c]{
Instituto de F\'{i}sica T\'{e}orica UAM/CSIC, 
calle Nicol\'{a}s Cabrera 13-15, Cantoblanco, 28049, Madrid, Spain}
\affiliation[d]{
Institute of Cosmology and Gravitation, University of Portsmouth\\
Dennis Sciama Building, Portsmouth PO1 3FX, United Kingdom}

\emailAdd{e.dimastrogiovanni@rug.nl}
\emailAdd{matteo.fasiello@csic.es}
\emailAdd{emir.gumrukcuoglu@port.ac.uk}

\abstract{We consider the possibility of extra spinning particles during inflation, focussing on the spin-2 case. Our analysis relies on the well-known fully non-linear formulation of interacting spin-2 theories. We explore the parameter space of the corresponding inflationary Lagrangian and identify regions therein exhibiting signatures within reach of upcoming CMB probes. We provide a thorough study of the early and late-time dynamics ensuring that stability conditions are met throughout the cosmic evolution. We characterise in particular the gravitational wave spectrum and three-point function finding a local-type non-Gaussianity whose amplitude may be within the sensitivity range of both the LiteBIRD and CMB-S4 experiments.

}

\begin{document}
\maketitle
\flushbottom

\section{Introduction}
\label{sec:intro}

The existence of an early phase of accelerated expansion, inflation, is by now a centrepiece of our understanding of how the universe begun \cite{Lyth:1998xn} and has come to assume its current structure. A scalar field with a sufficiently flat potential can drive inflation long enough to make sense of key observations, such as those of the cosmic microwave background (CMB) anisotropies. Within the single-field slow-roll paradigm, one may find several consistent realisations of a technically natural small inflaton mass (see e.g. \cite{Pajer:2013fsa}) and verify complete agreement with current observational bounds on inflation. These successes notwithstanding, ours remains a broad-brush picture of inflation. Key questions remain unanswered, including fundamental ones: 
what is the energy scale of inflation? what is the early phase field content?

 The available constraints on inflationary parameters such as the tensor-to-scalar ratio $r$ and non-Gaussianity $f_{\rm NL}$ leave the door open to a plethora of intriguing possibilities, an embarrassment of riches one would like to sacrifice in favour of more a well-defined inflationary dynamics. The unprecedented array of cosmological probes scheduled for  launch in the coming years promises to propel us in this very direction. The projected constraints on cosmological parameters will cross \textit{qualitative thresholds}, including those that will help us solve the single \textit{vs} multi-field dichotomy. One should mention here the target of $\sigma_r\sim 10^{-3}$ within reach for CMB-S4 experiments \cite{Abazajian:2016yjj} and the LiteBIRD satellite \cite{Lee2019LiteBIRD}, which will rule in (out) celebrated models such as Starobinsky inflation \cite{Starobinsky:1980te}. 
The constraints on (scalar) non-Gaussianity from Euclid \cite{Amendola:2012ys}, LSST \cite{Abell:2009aa} and SKA \cite{Camera:2014bwa,Bacon:2018dui} will be just as consequential: these may go as far as $\sigma_{f_{\rm NL}}\sim 1$ and down an order of magnitude or so \cite{Munoz:2015eqa} with 21cm cosmology. In typical single-field slow-roll realisations non-Gaussianities are much smaller than unity. On this basis, a near future detection of primordial non-Gaussianity may provide the ``smoking gun'' for multi-field inflation. 

Remarkably, the last decade has witnessed the rise of new probes of gravity. Following the first direct gravitational wave (GW) detection in 2015, these probes have led to spectacular new insights into gravitational wave astronomy and hold the potential to transform also our understanding of early universe physics. Ground (and, soon, space-) based interferometers access very small scales thus opening up, together with pulsar timing arrays, an observational window on inflation that is complementary to both CMB and large scale structure (LSS) measurements. For example, the detection of  a primordial GW signal at small scales would strongly point to multi-field inflationary mechanisms. Further characterisation of the GW signal may reveal much more specific information about inflation. A case in point is the possibility of a \textit{chiral} GW spectrum, often associated to Chern-Simons-type couplings \cite{Anber:2009ua,Bartolo:2017szm}. 

The opportunities afforded by a multi-probe and multi-scale search for imprints of early universe physics call for a full characterisation of the signatures associated to the inflationary particle content, going well beyond the single-field hypothesis. The reasons for doing so are manifold. From the top-down perspective, it is clear that UV finite theories such as string theory point to a rich inflationary field content \cite{Baumann:2014nda}. From moduli fields to gauge fields to Kaluza Klein modes, all can be 4-dimensional end products of higher dimensional constructions. A multi-field mechanism is not just a ``natural''  scenario but also an eminently testable one. Besides the discovery potential stemming from exploring primordial GW scale dependence and chirality, non-Gaussianity provides the most powerful handle on the inflationary particle content. From the momentum scaling, angular dependence, and possible periodic features of squeezed non-Gaussianities we can infer crucial information on the mass, spin, and couplings of the inflationary content \cite{Kehagias:2015jha,Arkani-Hamed:2015bza}.

There is a rich and interesting literature on the model-building aspects and the signatures of extra scalar and vector fields during inflation \cite{Martin:2013tda}. In this work we shall focus instead on the possibility of higher spin particles, and in particular spin-2 fields. Our starting point is the relatively recent formulation \cite{deRham:2010kj,Hassan:2011zd,Hinterbichler:2012cn} of a consistent fully non-linear theory of interacting spin two theories. These models have been widely used in attempts to deliver a technically natural mechanism to drive the universe late-time acceleration \cite{deRham:2014zqa}. In our present work, we adopt a different perspective, namely that of  studying the cosmology associated to the presence of an extra spin-2 field $f$ that couples to gravity (i.e. the usual massless spin-2 field $g$ that couples to matter) in a consistent manner. Crucially, already the results in \cite{Boulanger:2000rq} indicated that in a theory of interacting spin-2 particles there may be at most one massless spin-2 field. That place having been taken up by the massless field $g$, the theory with both the $g$ and $f$ field will necessarily contain massive eigenstates. 

We provide a thorough study of the cosmology of consistent interacting spin-2 theories in the inflationary context. The presence of different tensor sectors, such as those due to multiple spin-2 particles during inflation, offers the intriguing possibility of sourcing gravitational waves at linear order, in contradistinction to single-field scenarios. As we shall see, large GW non-Gaussianities are perhaps the most notable signature associated with the field content we explore in this manuscript. 

This paper is organised as follows. In \textit{Section} \ref{sec:bigravity} we study the background solution and the perturbative stability of our set-up. We provide extensive details for the benefit of the reader who may not be familiar with the intricacies stemming from requiring a stable FRW cosmology for a theory with multiple spin-2 fields.  \textit{Section} \ref{sec:spectra} is devoted to the calculation of the GW power spectrum and the three-point function amplitude and scale dependence. In \textit{Section} \ref{Allowed parameter space}  we identify the allowed parameter space of the model and provide in \textit{Section} \ref{discussion} a concrete example with corresponding signatures. In   
\textit{Section} \ref{sec:conclusions} we offer our conclusions and point to future work. For completeness, we report in  \textit{Appendix} \ref{app:allterms} the additional details necessary to reproduce our calculations. \\

\section{The theory of extra spin--2 field with gravitational interactions}
\label{sec:bigravity}

In this section we review the theoretical framework at the heart of our analysis and discuss the conditions that arise from a consistent cosmological evolution. We are after a theory of cosmic inflation driven by a scalar field but whose overall particle content is enriched by an extra spin-2 field. Fortunately, in describing such an early universe set-up, we can immediately benefit from an extensive literature on spin-2 particles at late times. The background being that of Friedmann-Lema\^itre-Robertson-Walker (FLRW) spacetime, we will be able to exploit here much of the late universe results and ``technology''.\\
We start with the notion that an additional interactive spin-2 particle in the inflationary context (i.e. with a scalar and a massless spin-2 field already in place) is necessarily a massive one \cite{Boulanger:2000rq}. The fully non-linear consistent  formulation of a spin-2 massive theory was first put forward in \cite{deRham:2010ik,deRham:2010kj}. In our set-up the fiducial metric of \cite{deRham:2010kj} becomes dynamical, making the Hassan-Rosen Lagrangian \cite{Hassan:2011zd} together with a  ``matter'' sector Lagrangian $\mathcal{L}_{\phi}$ that describes the slowly rolling inflaton our ideal starting point:
\begin{equation}
S = \int d^4x \left[\frac{M_g^2}{2} \sqrt{-g} \,R[g] + \frac{M_f^2}{2} \sqrt{-f} \, R[f] -m^2 M^2 \sqrt{-g}\sum_{n=0}^4 \beta_n\,e_n(\sqrt{g^{-1}f}) + \sqrt{-g}\,\mathcal{L}_{\phi}\right]\;.
\label{eq:initial-action}
\end{equation}
 In the above, the matter follows the geodesics of the tensor $g_{\mu\nu}$ which we thus interpret as the \emph{physical metric}. The second dynamical tensor field $f_{\mu\nu}$ represents the additional spin--2 field. The coupling between the two tensors $g$ and $f$ is given by the Lorentz invariant de Rham-Gabadadze-Tolley (dRGT) terms, which are constructed by the elementary polynomials $e_n$. These are defined via the determinant expansion $\det(\mathbb{1}+X) = \sum_{n=0}^4 e_n(X)$, as
\begin{align}
e_0(X) &\equiv1\,,\nonumber\\
e_1(X) &\equiv[X]\,,\nonumber\\
e_2(X) & \equiv\frac{1}{2!}\,\left([X]^2 -[X^2]\right)\,,\nonumber\\
e_3(X) & \equiv\frac{1}{3!}\,\left([X]^3 - 3\,[X]\,[X^2]+2\,[X^3]\right)\,,\nonumber\\
e_4(X) & \equiv\frac{1}{4!}\,\left([X]^4 +8\,[X]\,[X^3]-6\,[X]^2[X^2] +3\,[X^2]^2-6\,[X^4]\right)\,,
\end{align}
where square brackets around tensorial quantities represent the trace operation.
Following the convention of Ref.~\cite{Dimastrogiovanni:2018uqy}, we also define
\begin{equation}\label{Mdef}
\frac{1}{M^2} \equiv \frac{1}{M_g^2}+\frac{1}{M_f^2}\,,
\end{equation}
such that the two-tensor interaction term is symmetric under the $g\leftrightarrow f$ exchange.

\subsection{Review of background cosmology}
\label{sec:background}
For the background configuration, we aim at a flat FLRW cosmology in the $g$-metric. We also restrict the $f$--metric to be homogeneous and isotropic with a vanishing spatial curvature. The corresponding line elements are
\begin{align}
g_{\mu\nu}dx^\mu dx^\nu &=-N^2dt^2  + a^2 \delta_{ij} dx^idx^j\,,\nonumber\\
f_{\mu\nu}dx^\mu dx^\nu &=\xi^2 \left(-N^2\tilde{c}^2 dt^2  + a^2 \delta_{ij}dx^idx^j\right)\,,
\label{eq:background-metric}
\end{align}
where $\xi$ quantifies the ratio of the scale factors in the two metrics, while $\tilde{c}$ defines the light cone on the $f$--metric. In order to determine the background dynamics, we keep explicit the freedom on time reparametrization. We will later fix the cosmological time by setting $N=1$.

Combining the zero-order contributions from kinetic \eqref{eq:actionexpand-kinetic}, interaction \eqref{eq:actionexpand-interaction} and matter \eqref{eq:actionexpand-matter} terms, we can write down the total mini-superspace action as
\begin{equation}
S_{\rm mss} = V\,\int dt \,a^3N\left[
-\frac{3\,M_g^2\dot{a}^2}{a^2N^2}
-\frac{3\,M_f^2(\xi\,\dot{a}+a\,\dot{\xi})^2}{a^2\tilde{c}\,N^2}-m^2M^2(\rho_{m,g}+\tilde{c}\,\xi^4\rho_{m,f})+P
\right]\,,
\label{eq:minisuperspace}
\end{equation}
where $P$ represents the analogue pressure for the inflaton field. In the rest of the paper we use the following functions to represent convenient combinations of the $\beta_n$ parameters
\footnote{The definition of function $\Gamma$ in Ref.~\cite{DeFelice:2014nja} differs from the one here by an additional term $(\tilde{c}-1)Q$. The two definitions coincide in the limit $\tilde{c}=1$.}
\begin{align}
\rho_{m,g} \equiv&\beta_0+3\,\beta_1\xi+3\,\beta_2\xi^2+\beta_3\xi^3\,,
\nonumber\\
\rho_{m,f} \equiv & \beta_4+\frac{3\,\beta_3}{\xi}+\frac{3\,\beta_2}{\xi^2} + \frac{\beta_1}{\xi^3}\,,
\nonumber\\
\Gamma \equiv & \xi\,(\beta_1+2\,\beta_2\xi+\beta_3\xi^2)\,,
\nonumber\\
Q \equiv &\xi^2(\beta_2+\beta_3\xi)\,.
\label{eq:GammaQdefined}
\end{align}

Making use of Eqs.\eqref{eq:useful-matter}, we vary the mini-superspace action \eqref{eq:minisuperspace} with respect to $N$, $\tilde{c}$, $a$, $\xi$ and $\phi$. We obtain he following set of equations of motion: 
\begin{align}
%%%
3\,M_g^2H^2 &= m^2M^2\rho_{m,g}+\rho\,,
\label{eq:eqH}
\\
%%%
3\,M_f^2H_f^2 & =m^2M^2\rho_{m,f}\,,
\label{eq:eqHf}
\\
%%%
2\,M_g^2\,\dot{H} & =m^2M^2(\tilde{c}-1)\Gamma - (\rho+P)
\,,
\label{eq:eqHdot}
\\
%%%
2\,M_f^2\,\dot{H}_f&=-\frac{m^2M^2}{\xi^3}(\tilde{c}-1)\Gamma
\,,
\label{eq:eqHfdot}
\\
%%%
\dot{\rho}&=-3\,H\,(\rho+P)
\,,
\label{eq:eqrho}
\end{align}
where we have set $N=1$ and defined the expansion rates for the two metrics as
\begin{equation}
 H \equiv \frac{\dot{a}}{a}\,,\qquad
 H_f \equiv \frac{1}{\tilde{c}\,\xi}\,\left(H + \frac{\dot{\xi}}{\xi}\right)\,.
 \label{eq:HHfdefined}
\end{equation}
Additionally, the contracted Bianchi identity for either metric gives rise to the constraint:
\begin{equation}
\Gamma\,(H-H_f\xi)=0\,.
\label{eq:eqcons}
\end{equation}
This equation has two solutions. The first one, $\Gamma=0$, is known to lead to a non-linear ghost instability  \cite{Gumrukcuoglu:2011zh,DeFelice:2012mx}, a result that holds true also in theories with multiple spin-2 fields \cite{Comelli:2012db,DeFelice:2014nja}. We will then restrict our analysis to the case where 
\begin{equation}
H = H_f\,\xi\,.
\label{eq:constraint}
\end{equation}
The constraint implies two consistency relations. The first arises from requiring that \eqref{eq:eqHf} and \eqref{eq:eqH} are consistent under \eqref{eq:constraint}:
\begin{equation}
\rho_{m,g}-\frac{M_g^2\xi^2}{M_f^2}\,\rho_{m,f} = -\frac{\rho}{m^2M^2}\,,
\label{eq:eqX}
\end{equation}
which can be seen as the equation that determines $\xi$. The second one arises from the consistency of \eqref{eq:eqHfdot} with \eqref{eq:eqHdot},
\begin{equation}
2(\tilde{c}-1)W = \frac{P+\rho}{M_g^2}\,,
\label{eq:eqct}
\end{equation}
where we defined
\begin{equation}
W \equiv \frac{m^2(1+\kappa\,\xi^2)\,\Gamma}{2(1+\kappa)\xi^2}-H^2\,,
\label{eq:defW}
\end{equation}
and $\kappa\equiv M_f^2/M_g^2$.
The function $W$ is the {\it Higuchi function}, since $W>0$ is the dRGT generalisation of the Higuchi bound \cite{Fasiello:2012rw,Fasiello:2013woa,DeFelice:2014nja}.
The consistency equation \eqref{eq:eqct} is an algebraic equation that determines $\tilde{c}$. Using the background equations, we can write the solution for $\tilde{c}$ in terms of the equation of state of matter $w\equiv P/\rho$ and $\xi$ as
\begin{equation}
\tilde{c}-1 = (1+w)\,\frac{3\,\xi^2(\xi^2\rho_{m,f}-\kappa\,\rho_{m,g})}{3\,(1+\kappa\,\,\xi^2)\Gamma-2\,\xi^4\rho_{m,f}}\,.
\label{eq:eqct2}
\end{equation}
Finally, using \eqref{eq:HHfdefined} and \eqref{eq:constraint}, we can express the time variation of $\xi$ in terms of these functions via
\begin{equation}
\frac{\dot{\xi}}{H\,\xi}= \tilde{c}-1 = (1+w)\,\frac{3\,\xi^2(\xi^2\rho_{m,f}-\kappa\,\rho_{m,g})}{3\,(1+\kappa\,\xi^2)\Gamma-2\,\xi^4\rho_{m,f}}\,.
\label{eq:eqdotxi}
\end{equation}

At this point, we can in principle replace $\xi$ and $\tilde{c}$ by solving Eqs.\eqref{eq:eqX} and \eqref{eq:eqct} to remove all references to the metric $f$. However, \eqref{eq:eqX} is a fourth order algebraic equation for $\xi$ and different solutions have different implications for the cosmological evolution. The requirement of a sensible background evolution which is stable against small perturbations is known to severely limit the cosmological solution in the late Universe \cite{Comelli:2011zm, Comelli:2012db, DeFelice:2014nja}. In the following section, we revisit these requirements by focussing on an inflaton dominated universe.

\subsection{Perturbative stability}
\label{sec:stability-conditions}
For a given dynamical degree of freedom, the free Lagrangian in Fourier space, quadratic in the field, is formally 
\begin{equation}
\mathcal{L}_k \ni \mathcal{K}_\psi\, \left(|\dot{\psi}_k|^2 -(c_\psi^2\,k^2+m_\psi^2)|\psi|^2\right)\,.
\label{eq:formal-lagrangian}
\end{equation}
For small perturbations, there are in general three stability requirements, which protect the background from ghost, gradient and tachyonic instabilities. In \eqref{eq:formal-lagrangian}, these  correspond respectively to a positive kinetic term $\mathcal{K}_\psi>0$, a real propagation speed $c_\psi^2>0$ and a real mass $m_\psi^2>0$. 

In a cosmology with multiple spin-2 fields, the dynamical metric $f_{\mu\nu}$ brings in a new tensor perturbation adding two tensorial degrees of freedom. However, due to the two--metric interaction term breaking the two diffeomorphism symmetries down to one, a vector (i.e. two additional degrees of freedom) and a scalar mode arise in the gravitational sector. Moreover, we are including a scalar field in the matter sector. As a result, the dynamical and independent degrees of freedom comprise 4 tensor, 2 vector and 2 scalar perturbations. Each of these degrees of freedom need to pass the perturbative stability test around the cosmological background. 

One challenge is that both the tensor and scalar sectors in our setup are coupled systems with time-dependent coefficients. In particular, it is not possible to exactly diagonalise the Lagrangian for the scalar perturbations to obtain the mass eigenstates, although the kinetic terms and propagation speeds can be determined unambiguously. Fortunately, the instabilities generated by the imaginary part of the eigenmasses would be mild in our context, therefore we will disregard the conditions arising from avoiding tachyonic instabilities \footnote{The mass of gravitational perturbations are typically of order $H$ or smaller. Therefore these instabilities would take longer than the age of the universe to develop.}.
For the remaining instabilities, we adapt the conditions to avoid ghost and gradient instabilities in all sectors that have been obtained previously in Ref.\cite{DeFelice:2014nja} 
\footnote{The main difference with Ref.\cite{DeFelice:2014nja} arises from the normalisation of the interaction term, which can be accommodated by rescaling $m^2 \to m^2\kappa/(1+\kappa)$ when changing notations.
}.

For the tensor perturbations, only the kinetic term for the $f$--metric perturbations brings a stability condition (see Eq.\eqref{eq:action-quadratic})
\begin{equation}
\mathcal{K}_{\rm tensor}\propto \tilde{c} >0 \,.
\label{eq:condition-tensor}
\end{equation}

For the vector modes, positivity of the kinetic term in the UV and reality of propagation speed impose \cite{DeFelice:2014nja}
\begin{equation}
\mathcal{K}_{\rm vector}\propto \frac{\Gamma}{\tilde{c}+1} >0\,,\qquad
c^2_{\rm vector} \equiv 
1+\frac{P+\rho}{4\,M_g^2W}+\frac{Q\,(P+\rho)}{2\,M_g^2W\,\Gamma}\,\left(1+\frac{P+\rho}{4\,M_g^2W}\right)>0\,.
\label{eq:condition-vector}
\end{equation}
Finally for the scalar modes, avoiding a ghost in the UV demands that \cite{DeFelice:2014nja}
\begin{equation}
\mathcal{K}_{\rm matter} \propto \frac{\rho+P}{c_{s}^2}>0\,,
\qquad
\mathcal{K}_{\rm scalar} \propto \Gamma\,W >0\,,
\label{eq:condition-scalar1}
\end{equation}
where $c_s^2\equiv dP/d\rho$ for a scalar field is defined in \eqref{eq:density-soundspeed}. For stable gradients, the reality of propagation speeds leads to \cite{DeFelice:2014nja}
\begin{equation}
c_{\rm matter}^2 = c_{s}^2 >0\,,
\qquad
c_{\rm scalar}^2\equiv
1+\frac{2\,Q\,(P+\rho)}{3\,M_g^2 W\,\Gamma}+\frac{H^2(2\,Q-\Gamma)(P+\rho)}{6\,M_g^2W^2\Gamma}>0\,.
\label{eq:condition-scalar2}
\end{equation}

Thus, throughout the evolution, we need to make sure that the conditions \eqref{eq:condition-tensor}--\eqref{eq:condition-scalar2} are not violated. In the present study, the matter sector consists of a scalar field with a canonical kinetic term, rolling down a positive potential. As a result $c_s^2=1$ and $\rho+P>0$ are automatically satisfied, while the equation of state is bound to the range $-1\leq w\leq 1$ by construction. In this case, Eq.\eqref{eq:eqct} implies that as long as $W>0$ is satisfied, we have $\tilde{c}>1$, automatically fulfilling the requirement for the stability of tensor perturbations \eqref{eq:condition-tensor}. Moreover, the propagation speed for the vector mode can be written as
\begin{equation}
 c_{\rm vector}^2 = \frac{W}{2}\,\left(1+\frac{P+\rho}{4\,M_g^2W}\right)\left[
 \frac{1+3\,c_{\rm scalar}^2}{H^2+2\,W}+\frac{2\,H^2}{W\,(H^2+2\,W)}\,\left(1+\frac{P+\rho}{4\,M_g^2W}\right)
 \right]\,,
\end{equation}
which is manifestly positive if the conditions $W>0$, $P+\rho>0$ and $c_{\rm scalar}^2>0$ are satisfied. 

In summary, for a matter sector made up of a canonical scalar with a positive potential, the independent stability conditions that are required to be true throughout the evolution are
\begin{equation}
\Gamma>0\,,\quad
W>0\,,\quad
c_{\rm scalar}^2 > 0\,.
\label{eq:stability}
\end{equation}

\subsection{Branching of the evolution}
\label{sec:branches}
The equation for $\xi$ \eqref{eq:eqX}, in terms of the $\beta_i$, can be written as
\begin{equation}
\frac{\beta_1}{\kappa}\,\frac{1}{\xi} + \left(-\beta_0+\frac{3\,\beta_2}{\kappa}\right)+3\,\left(-\beta_1+\frac{\beta_3}{\kappa}\right)\,\xi+\left(-3\,\beta_2+\frac{\beta_4}{\kappa}\right)\,\xi^2-\beta_3\xi^3 = \frac{1+\kappa}{\kappa}\,\frac{\rho}{m^2M_g^2}\,.
\label{eq:eqXbetas}
\end{equation}
In general, this equation has four roots. We find it useful to discuss some properties of the $\xi$ solution in two distinct asymptotic limits that 
depend on the ratio $\rho/(m^2M_g^2)$. In the late time asymptotics, we are in the {\it low energy} regime, where $\rho\ll m^2 M_g^2$ and $\xi$ tends to a constant value $\xi\to \xi_c$. Conversely, in the early time asymptotics, we are in the {\it high energy} regime such that $\rho\gg m^2 M_g^2$. 

As first shown in Ref.\cite{Comelli:2011zm}, the stability of the background solution in the high energy regime is highly sensitive to the specific evolution of $\xi$ as well as the equation of state of the dominant matter fluid. In this section, we catalogue the various evolution branches and determine how the stability conditions manifest themselves in each case. Although the early behaviour poses a roadblock to many dark energy applications \cite{Comelli:2012db,Comelli:2014bqa,Konnig:2014xva,Kenna-Allison:2018izo}, we show that there is one evolution branch that leads to a healthy solution in the context of inflation. We then review the late time attractor solution where we can interpret the second metric as an external spin--2 field on top of standard inflationary universe.

\subsubsection{Early times (high energy)}
\label{sec:high-energy}
The different branches arise from the early asymptotics $\rho\gg m^2M_g^2$. As we go back in time, the right hand side of \eqref{eq:eqXbetas} turns on, thus the left hand side should grow accordingly. 
We then define two main branches in this case:
\begin{itemize}
 \item {\bf Branch A:}  $\lim_{a\to 0}\xi = 0$, i.e. $\xi(t) <\xi_c$ with $\dot \xi>0$;
 \item {\bf Branch B:}  $\lim_{a\to 0}\xi =\infty$, i.e. $\xi(t) >\xi_c$ with $\dot \xi<0$.
\end{itemize}
At this point, we already have enough information about Branch B to determine its stability properties. Using Eq.\eqref{eq:eqct}, we replace $\tilde{c}$ in Eq.\eqref{eq:eqdotxi} to obtain
\begin{equation}
\dot \xi= \frac{H\,\xi\, (P+\rho)}{2\,W\,M_g^2}\,.
\label{eq:dotxiW}
\end{equation}
For the matter source that satisfies the null energy condition, which is the case for the scalar inflaton in our construction, we see that $\dot\xi<0$ implies $W<0$. In other words, Branch B suffers from the Higuchi instability. We therefore focus only on Branch A for the discussion of the high energy limit.

In Branch A, $\xi^{-1}$ term dominates the Eq.\eqref{eq:eqXbetas} at early times. Keeping up to next-to-leading order terms, we have
\begin{equation}
\xi^A=\frac{\beta_1}{1+\kappa}\,\frac{m^2M_g^2}{\rho}\left[1+ \frac{m^2M_g^2(3\,\beta_2-\kappa\,\beta_0)}{(1+\kappa)\rho}+\mathcal{O}
\left(\frac{m^2M_g^2}{\rho}\right)^{2}\right]\,,
\end{equation}
where the positivity of $\xi$ requires
\begin{equation}
 \beta_1>0\,.
 \label{eq:stability-early}
\end{equation}
Thus, for negative $\beta_1$, this branch would not give a physical solution and only the unstable Branch B solutions would be available.

The remaining equations of motion, when expanded for large density, yield
\begin{align}
\tilde{c}-1 =& 3\,(1+w)\left[1 + \frac{m^2M_g^2(3\,\beta_2-\kappa\,\beta_0)}{(1+\kappa)\rho}+\mathcal{O}
\left(\frac{m^2M_g^2}{\rho}\right)^{2}
\right]
\,,\nonumber\\
3\,H^2 =&
\frac{\rho}{M_g^2}  \left[1 + \frac{m^2M_g^2\kappa\,\beta_0}{(1+\kappa)\rho}
+\mathcal{O}
\left(\frac{m^2M_g^2}{\rho}\right)^{2}
\right]
\,,\nonumber\\
2\,\dot{H} =&
-\frac{\rho+P}{M_g^2}
\left[1-\frac{3\,m^4\,M_g^4\kappa\,\beta_1^2}{(1+\kappa)^2\rho^2}
+\mathcal{O}
\left(\frac{m^2M_g^2}{\rho}\right)^{3}
\right]\,.
\end{align}

In this limit, the key quantities that appear in the stability conditions \eqref{eq:stability} become
\begin{align}
\Gamma =& \frac{m^2\,M_g^2\,\beta_1^2}{(\kappa+1)\,\rho} + \mathcal{O}
\left(\frac{m^2M_g^2}{\rho}\right)^{2}\,,
\nonumber\\
W =&
\frac{\rho}{6\,M_g^2} + m^2\mathcal{O}(1)\,,
% \,,
\nonumber\\
c^{2}_{\rm scalar} =&
-(1+2\,w)
+\mathcal{O}
\left(\frac{m^2M_g^2}{\rho}\right)\,.
\end{align}
The condition $W>0$ is automatically satisfied for this branch, as we predicted in Eq.\eqref{eq:dotxiW}. Combined with the matter stability condition $|w|<1$, the range for the equation of state that keeps $c^2_{\rm scalar}$ positive is
\begin{equation}
 -1<w<-\frac12\,.
\end{equation}

This branch has sensible Friedmann equations and has no early-time ghost instability. Ref.\cite{Comelli:2012db} found that this branch contains a gradient instability in the dark energy context, since for radiation and matter dominated universe, one has $c_{\rm scalar}^2<0$ at early times.\footnote{This instability may be tamed by non-linear terms \cite{Aoki:2015xqa,Mortsell:2015exa,Hogas:2019ywm}.} This conclusion of course applies only to fluids with $w>-1/2$. However, in the context of inflation, we can assume that the beginning of inflation is extended to the energy scales that correspond to the cut-off of the effective theory. Provided that the transition to the low energy regime occurs while inflation is in progress, i.e. $w\sim -1$, this instability is never invoked.
 \subsubsection{Late times (low energy)}
\label{sec:low-energy}
At late times, the right hand side of \eqref{eq:eqXbetas} decreases and the solution for $\xi$ converges to a constant $\xi=\xi_c$ defined by
\begin{equation}
\frac{\beta_1}{\kappa}\,\frac{1}{\xi} + \left(-\beta_0+\frac{3\,\beta_2}{\kappa}\right)+3\,\left(-\beta_1+\frac{\beta_3}{\kappa}\right)\,\xi+\left(-3\,\beta_2+\frac{\beta_4}{\kappa}\right)\,\xi^2-\beta_3\xi^3\Bigg\vert_{\xi=\xi_c} =0\,.
\label{eq:eqXbetas-late}
\end{equation}
Although there are four different end-points for the $\xi$ evolution, the late time approach to these values is universal. Defining
\begin{equation}
\rho_{m,g~ c}  = \rho_{m,g}\vert_{\xi=\xi_c}\,,\qquad
\rho_{m,f~ c}  = \rho_{m,f}\vert_{\xi=\xi_c}\,,\qquad
\Gamma_{c}  = \Gamma\vert_{\xi=\xi_c}\,,\qquad
Q_c = Q\vert_{\xi=\xi_c}\,,
\end{equation}
and expanding for $\rho\ll m^2M_g^2$
\begin{align}
\xi = & \xi_c \left[1-\left(\frac{(1+\kappa)\xi_c^2}{3\,(1+\kappa\,\xi_c^2)\,\Gamma_c-2\,\kappa\,\xi_c^2\,\rho_{m,g~c}}\right)
\frac{\rho}{m^2M_g^2}+ \mathcal{O}\left(\frac{\rho}{m^2M_g^2}\right)^2\right]\,,
\nonumber\\
\tilde{c} =& 1+ \frac{\rho+P}{m^2M_g^2} \left[
\frac{3(1+\kappa)\xi_c^2}{3\,(1+\kappa\,X_\xi^2)\,\Gamma_c-2\,\kappa\,\xi_c^2\,\rho_{m,g~c}}
+ \mathcal{O}\left(\frac{\rho}{m^2M_g^2}\right)
\right]\,.
\end{align}
The equations of motion for the $g$--metric become
\begin{align}
3\,H^2 = &
\frac{m^2\kappa\,\rho_{m,g~c}}{1+\kappa} +\frac{(3\,\Gamma_c-2\,\kappa\,\xi_c^2\rho_{m,g~c})}{3\,(1+\kappa\,\xi_c^2)\,\Gamma_c-2\,\kappa\,\xi_c^2\,\rho_{m,g~c}}\,\frac{\rho}{M_g^2}+m^2\,\mathcal{O}\left(\frac{\rho}{m^2M_g^2}\right)^2
\,,\nonumber\\
2\,\dot{H} =& -\frac{(3\,\Gamma_c-2\,\kappa\,\xi_c^2\rho_{m,g~c})}{3\,(1+\kappa\,\xi_c^2)\,\Gamma_c-2\,\kappa\,\xi_c^2\,\rho_{m,g~c}}\,\frac{\rho+P}{M_g^2}
+m^2\,\mathcal{O}\left(\frac{\rho}{m^2M_g^2}\right)^2\,.
\label{eq:Friedmann-late}
\end{align}
At late times, some of the functions in the stability conditions \eqref{eq:stability} are
\begin{align}
W =& \frac{m^2}{1+\kappa}\left(
\frac{\Gamma_c(1+\kappa\,\xi_c^2)}{2\,\xi_c^2}-\frac{\kappa\,\rho_{m,g~c}}{3}
\right)+m^2\,\mathcal{O}\left(\frac{\rho}{m^2M_g^2}\right)\,,
\nonumber\\
c^2_{{\rm scalar}} =& 1+ \mathcal{O}\left(\frac{\rho}{m^2M_g^2}\right)\,.
\end{align}
We see that the gradient instability in the scalar sector never appears, while the no-ghost conditions $W>0$ and $\Gamma>0$ are parameter dependent and can be satisfied by appropriate choices of $\Gamma_c$ and $\rho_{m,g~c}$. 
% %  

In the remainder of the paper, we consider a scenario where the transition from the high to low energy regime occurs during inflation, while the largest observable scale is deep inside the horizon. In the calculations of the correlators, we will therefore take
\begin{equation}
\xi \simeq \xi_c\,,\qquad
\tilde{c} \simeq 1\,.
\end{equation}
The second of these is automatic, while the specific value of $\xi_c$ can be chosen by fixing one of the $\beta_n$ parameters.

All late-time instabilities are avoided if
 \begin{equation}
 \Gamma_c> 0\,, \qquad
\frac{\Gamma_c(1+\kappa\,\xi_c^2)}{2\,\xi_c^2}-\frac{\kappa\,\rho_{m,g~c}}{3}>0\,.
\label{eq:stability-late}
 \end{equation}
It is important to note that in this low energy regime with $\rho\ll m^2 M_g^2$, the energy contribution from the 2--metric interaction term can be parametrically larger than the inflaton density, which may lead to de Sitter evolution to that goes on indefinitely. To avoid this, we set to zero the contribution to Friedmann equation:
\begin{equation}
\rho_{m,g~c} = 0\,.
\label{eq:impose-nocc}
\end{equation}
Evaluating Eq.\eqref{eq:eqX} in the late time asymptotics, the above selection implies that we also need to fix $\rho_{m,f~c} =0$. Both of these can be realised at the price of fixing two of the $\beta_n$ parameters. As a result of this choice, the background equations at late times \eqref{eq:Friedmann-late} become:
\begin{align}
3\,H^2\simeq &
\frac{\rho}{M_{eff}^2}
\,,\nonumber\\
2\,\dot{H}\simeq & -\frac{\rho+P}{M_{eff}^2}\,.
\label{eq:Friedmann-late-final}
\end{align}
Thus at the background level, the two metrics are decoupled and the resulting cosmology is simply the usual inflationary universe in General Relativity, with an effective Planck mass defined as
\begin{equation}
 M_{eff}^2 \equiv (1+\kappa\,\xi_c^2)\,M_g^2\,.
 \label{eq:Meff}
\end{equation}
This construction allows us to interpret the second metric as an external spin--2 field. In this case, all of the stability conditions in Eq.~\eqref{eq:stability} are satisfied at late times provided that $\Gamma_c>0$.% 
\subsection{Action for tensors in the low energy regime}
In this Section, we present the action up to cubic order in the low energy regime discussed in Sec.~\ref{sec:low-energy}. Excluding the background portion, we decompose the action as 
\begin{equation}
S= \int d^4x \,a^3 \left(
\mathcal{L}_{\rm tot}^{(2)}+\mathcal{L}_{\rm tot}^{(3)}+\dots
\right)\,.
\end{equation}
%
% %
The full expressions are given in Appendix~\ref{app:allterms}. Here, we present them in the low energy limit by fixing $\xi=\xi_c$ and $\tilde{c}=1$. From here on, we suppress the subscript $c$ that represents the constant values reached at late times. 

At quadratic order in tensor perturbations, the dynamics of the free tensor modes are described by \eqref{eq:action-quadratic}

\begin{align}\label{eqquad}
{\mathcal L} ^{(2)}=& \frac{M_g^2}{8}\,\Bigg[\dot{h}_{ij}\dot{h}^{ij}-\frac{1}{a^2}\partial_kh^{ij}\partial^k h_{ij}
+\kappa\,\xi^2\,\left(\dot{\gamma}_{ij}\dot{\gamma}^{ij}-\frac{1}{a^2}\partial_k\gamma^{ij}\partial^k \gamma_{ij}\right)
\nonumber\\
&
\qquad\qquad\qquad\qquad-\frac{m^2 \kappa\, \Gamma}{1+\kappa} \,(h_{ij}h^{ij}-2\,h^{ij}\gamma_{ij}+\gamma^{ij}\gamma_{ij})\Bigg]\,,
\end{align}
where we defined $\kappa \equiv M_f^2/M_g^2$.
% %
% %
We see that there is a constant interaction term proportional to $\Gamma$.\\
The cubic interactions that involve the tensor modes are \eqref{eq:action-cubic}
\begin{align}
{\mathcal L} ^{(3)}_{\rm tot} = &
-\frac{M_g^2}{4} h_{ij}\left[
\dot{h}^{ik}\dot{h}_k^{\;\;j}-\frac{1}{2\,a^2}
\Big(
 \partial^{i}h^{kl} \partial^{j}h^{}{}_{kl} 
- 2 \, \partial^{i}h_{kl} \partial^{l}h^{jk}
+ 2 \, \partial_{l}h^{i}_{\;\;k} \, \partial^{l}h^{jk} 
\Big)
\right]
\nonumber\\
&-\frac{M_f^2}{4}\xi^2\,\gamma_{ij}\left[
\dot{\gamma}^{ik}\dot{\gamma}_k^{\;\;j}-\frac{1}{2\,a^2}
\Big(
 \partial^{i}\gamma^{kl} \partial^{j}\gamma^{}{}_{kl} 
- 2 \, \partial^{i}\gamma_{kl} \partial^{l}\gamma^{jk}
+ 2 \, \partial_{l}\gamma^{i}_{\;\;k} \, \partial^{l}\gamma^{jk} 
\Big)
\right]
\nonumber\\
&
+\frac{m^2M^2}{48}\left(\Gamma-2\,Q\right)[(h-\gamma)^3]
+\frac{m^2M^2}{8}\Gamma\,[(h-\gamma)^2(h+\gamma)]\,.
\label{eq:action-cubic-late-time}
\end{align}
The non-derivative cubic interactions are controlled by combinations of $Q$ and $\Gamma$.

In Sec.\ref{sec:spectra}, we will be using these expressions to calculate the two-point and three-point correlation functions.

\section{Correlators amplitude and shape}
\label{sec:spectra}

\noindent We now proceed with the computation of the power spectrum and bispectrum for $h$, given by:
\begin{eqnarray}
	&&\langle h^{\lambda_{1}}_{\textbf{k}_{1}}h^{\lambda_{2}}_{\textbf{k}_{2}}\rangle=(2\pi)^{3}\delta^{(3)}(\textbf{k}_{1}+\textbf{k}_{2})\delta_{\lambda_{1}\lambda_{2}}P_{h}(k)\,,\\
	&&\langle h^{\lambda_{1}}_{\textbf{k}_{1}}h^{\lambda_{2}}_{\textbf{k}_{2}}h^{\lambda_{3}}_{\textbf{k}_{3}}\rangle=(2\pi)^{3}\delta^{(3)}(\textbf{k}_{1}+\textbf{k}_{2}+\textbf{k}_{3})B_{h}^{\lambda_{1}\lambda_{2}\lambda_{3}}(\textbf{k}_{1},\textbf{k}_{2},\textbf{k}_{3})\,.
\end{eqnarray}

\subsection{Power spectrum}
\label{sec:p-spectra}

The starting point for the power spectrum computation is the Lagrangian given in Eq.~(\ref{eqquad}). From here one obtains the equations of motion for the canonically normalised fields, $h_{c}\equiv a M_{g}h$ and $\gamma_{c}\equiv a M_{f}\xi \gamma$,
\begin{eqnarray}\label{eom2}
	&&h^{''}_{c}+\left(k^{2}+a^{2}m_h^2 -\frac{a^{''}}{a}\right)h^{}_{c}=a^{2}m_X^2 \gamma_{c}\,,\\
	&&	\gamma^{''}_{c}+\left(k^{2}+a^{2}m_\gamma^2-\frac{z^{''}}{z}\right)\gamma^{}_{c}=a^{2} m_X^2 h_{c}\,.
\end{eqnarray}
where $s\equiv a M_{g}$ and $z\equiv aM_{f}\xi$. Using Eq.~(\ref{Mdef}) and the definition of $\kappa$, the masses of the two modes are found to be equal to
\begin{equation}
	m_h^2 \equiv \frac{m^2\kappa\,\Gamma}{\kappa+1}\,,
	\qquad
	m_\gamma^2 \equiv \frac{m_h^2 }{\kappa\,\xi^2}\,,
	\label{eq:masshg}
\end{equation}
whereas the ``mass'' scale for the cross term is given by
\begin{equation}
	m_X^2 \equiv \frac{m_h^2 }{\sqrt{\kappa}\,\xi}\,.
	\label{eq:massx}
\end{equation}
The mixing term can be treated as a perturbation on top of the free Lagrangian as long as the following condition is in place
\begin{equation}
	\frac{m^2\,\sqrt{\kappa}\,\Gamma}{H^2(1+\kappa)\,\xi} \ll 1
	\label{eq:condition-I}
\end{equation}
or, equivalently, $m_{h}^{2}/H^{2}\ll \sqrt{\kappa}\xi$. From now on we will assume condition (\ref{eq:condition-I}), as this allows us to use the standard mode functions for free fields and introduce the small interaction via the \textsl{in-in} formalism. In addition, we must take the Higuchi bound into account. From the definition in \eqref{eq:defW}, the latter reads
\begin{equation}
	\frac{m^2 (1+\kappa\,\xi^2)\Gamma}{(1+\kappa)\,\xi^2} > 2\,H^2
	\label{eq:condition-IV}
\end{equation}
or, equivalently, as $m^{2}_{h}/H^{2}>2\kappa \xi^{2}/(1+\kappa \xi^{2})$, which is compatible with Eq.(\ref{eq:condition-I}) for 
\begin{equation}
	\sqrt{\kappa}\,\xi \ll 1\,.
	\label{eq:condition-III}
\end{equation}
In this regime, the three scales $m_h^2$, $m_X^2$ and $m_\gamma^2$ are organised according to the following hierarchy
\begin{equation}
	m_h^2 \ll m_X^2 \ll m_\gamma^2\,.
	\label{eq:hierarchy}
\end{equation}
Notice also that, for $\sqrt{\kappa}\xi\ll1$, Eq.~(\ref{eq:condition-I}) ensures that $h$ is fairly light. As one would expect (see also our explicit result in Eq.~(\ref{s-index})), this is crucial for the mode that couples to matter fields not to be suppressed at late times. \\

\begin{figure}[t]
	\centering
	\includegraphics[scale=0.22]{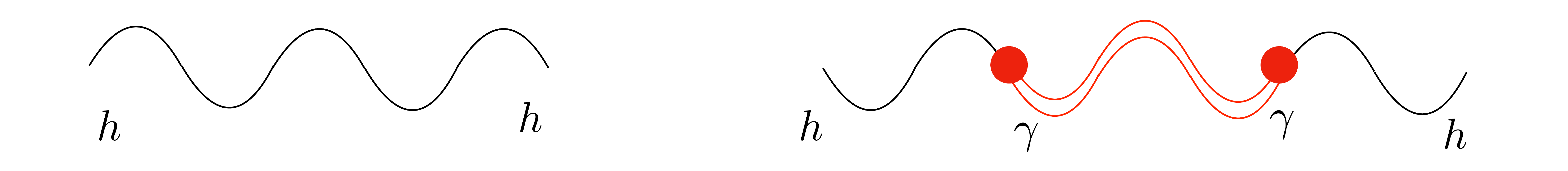}
	\caption[]{Left: the standard tensor two-point function from vacuum fluctuations. Right: the diagram corresponding to the power spectrum being ``sourced'' by the $h -\gamma$ mixing.}
	\label{fig:PS_diag}
\end{figure}

\noindent As we will explicitly verify below, the leading order contribution to $P_{h}$ is given by  the diagram on the left-hand side of Fig.~\ref{fig:PS_diag}. This is expected given that the 2-vertex interaction is taken to be small by construction.
Let us first derive the mode-function for $h$ from its free quadratic Lagrangian. Working at next-to-leading order in slow roll, the equation of motion reads
\begin{eqnarray}\label{eomh}
	&& h_{k}^{''}-2\left(\frac{1+\epsilon_{*}}{\tau}\right)h_{k}^{'}+\left(k^{2}+\frac{m_{h}^{2}}{H^{2}_{*}\tau^{2}} \right)h_{k}=0\,,
\end{eqnarray}
where the suffix ``*'' denotes a quantity calculated at the reference time $\tau_{*}$\footnote{Such reference point is typically chosen to be the (conformal) time at which the longest mode that is observationally relevant today exits the horizon.}. 
\noindent The solution for $h$ reads
\begin{equation}\label{solh}
	h_{k}(\tau)=(-\tau)^{3/2+\epsilon_{*}}\cdot\,\mathcal{A}\cdot\,H^{(1)}_{\nu_{T}}(-k\tau)\,,
\end{equation}
where $\nu_{T}\equiv\sqrt{9/4+3\epsilon-m_{h}^{2}/H_{*}^{2}}\simeq 3/2+\epsilon-m_{h}^{2}/(3H_{*}^{2})$ and
\begin{equation}
	\mathcal{A}\equiv \frac{H_{*}(1-\epsilon_{*})}{M_{g}}\sqrt\pi(-\tau_{*})^{-\epsilon_{*}} e^{i\left(\nu_{T}\cdot\pi/2+\pi/4 \right)}\,.
\end{equation}
\noindent  The mode-function in Eq.~(\ref{solh}) may be expanded at late times to give
\begin{eqnarray}
	\label{s-index}
	h_{k}(\tau)|_{k\tau\rightarrow 0}&= \frac{\sqrt{2} \,H^{}_{*}}{M_{g}^{}k^{3/2}}(-i)\cdot e^{i\left(\nu_{T}\cdot\pi/2+\pi/4 \right)}\cdot(-k\tau_{*})^{3/2-\nu_{T}}\cdot\left(\frac{\tau}{\tau_{*}} \right)^{\frac{m_{h}^{2}}{3H^{2}_{*}}}\,.
\end{eqnarray}
The corresponding contribution to the power spectrum then amounts to
\begin{equation}\label{p0}
	P_{h}^{(0)}= \frac{2\,H^{2}_{*}}{M_{g}^{2}k^{3}}\cdot(-k\tau_{*})^{3-2\nu_{T}}\cdot\left(\frac{\tau}{\tau_{*}} \right)^{\frac{2m_{h}^{2}}{3H^{2}_{*}}}\,.
\end{equation}
In this expression, $\tau/\tau_{*}$ corresponds to the number of e-folds, $N_{*}$, between the horizon exit of the mode $k_{*}\equiv -1/\tau_{*}$ and the end of inflation.\\

\noindent  Let us now compute the other contribution to the tree-level two-point function (depicted in the right panel of Fig.~\ref{fig:PS_diag}), which we will indicate as $P_{h}^{\text{mix}}$, 
\begin{eqnarray}
	P_{h}^{\text{mix}}=A^{\text{mix}}+B^{\text{mix}}\,,
\end{eqnarray}
where
\begin{eqnarray}
	&&A^{\text{mix}}\equiv \frac{\left(m^{2}M^{2}\Gamma\right)^{2}}{32}h_{k_{1}}(\tau) h_{k_{1}}^{*}(\tau)\Big\|    \int_{-\infty}^{\tau}d\tau_{1} \,a^{4}(\tau_{1})h_{k_{1}}(\tau_{1}) \gamma_{k_{1}}(\tau_{1})\Big\|^{2}\,,\\
	&&B^{\text{mix}}\equiv -\frac{\left(m^{2}M^{2}\Gamma\right)^{2}}{16}\text{Re}\left[h_{k_{1}}^{2}(\tau) \int_{-\infty}^{\tau}d\tau_{1} \,a^{4}(\tau_{1})h^{*}_{k_{1}}(\tau_{1}) \gamma_{k_{1}}(\tau_{1}) \int_{-\infty}^{\tau_{1}}d\tau_{2} \,a^{4}(\tau_{2})h_{k_{1}}^{*}(\tau_{2}) \gamma_{k_{1}}^{*}(\tau_{2}) \right] \,.\nonumber\\
\end{eqnarray}
\begin{figure}[t]
	\centering
	\includegraphics[scale=0.23]{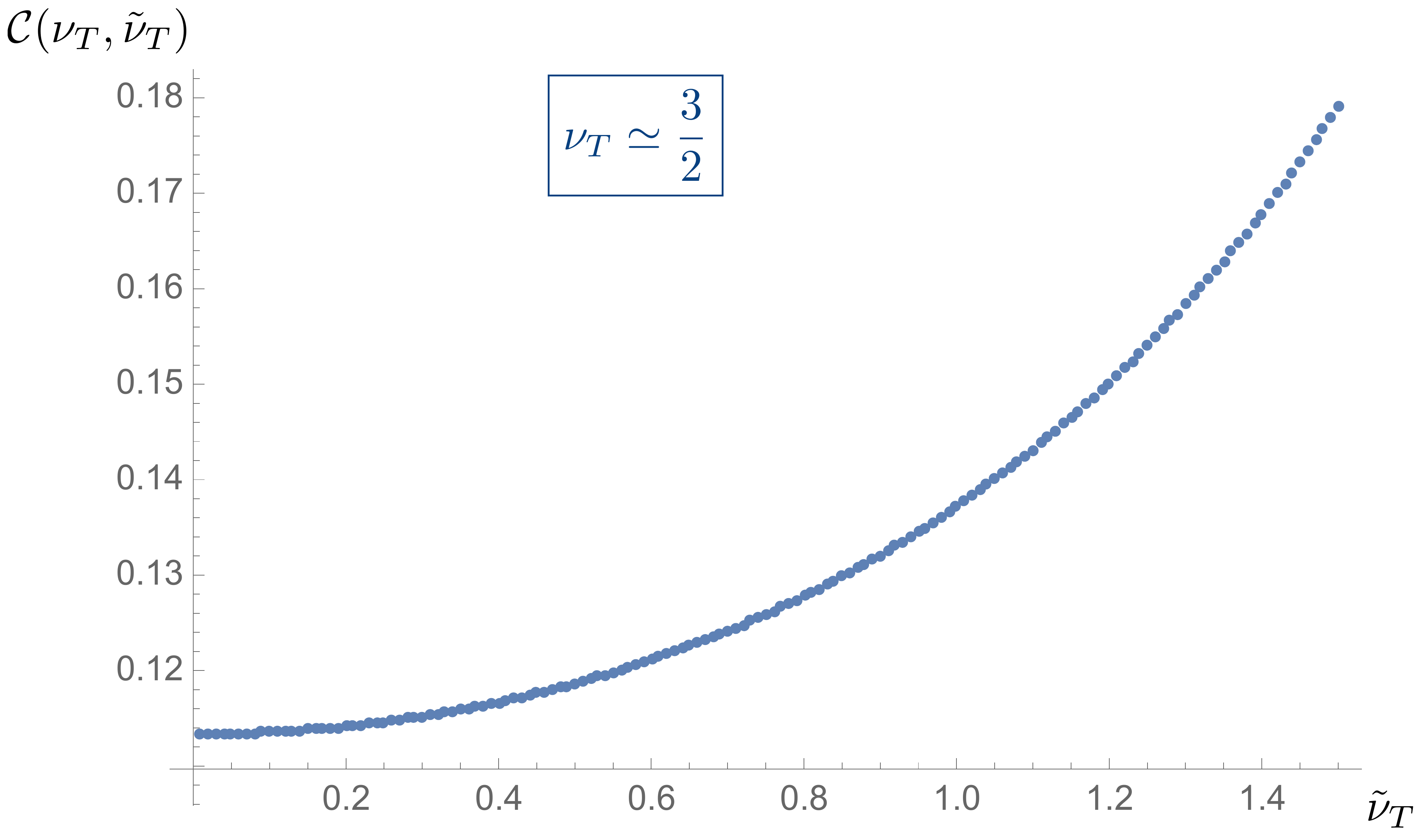}
	\caption[]{Plot of $\mathcal{C}(\nu_{T},\tilde{\nu}_{T})$ as a function of $\tilde{\nu}_{T}$, for small values of $m_{h}/H$. We have verified that, for $m_{h}^{2}/H^{2}\lesssim 5\times 10^{-2}$, $\mathcal{C}$ does not appreciably vary with $\nu_{T}$.}
	\label{cnunut}
\end{figure}
The mode function of $\gamma$ is obtained from the free-field equation of motion derived from Eq.~(\ref{eqquad}) and reads, this time to leading order in slow-roll,
\begin{equation}
	\gamma_{k}(\tau)=(-\tau)^{3/2}\cdot\,\frac{H_{}}{M_{g}\sqrt{\kappa}\xi}\sqrt{\pi}\,e^{i\left(\tilde{\nu_{T}}\cdot\pi/2+\pi/4 \right)}\cdot\,H^{(1)}_{\tilde{\nu}_{T}}(-k\tau)\,,
\end{equation}
where $\tilde{\nu}_{T}\equiv\sqrt{9/4-m_{\gamma}^{2}/H^{2}_{}}$, and we assumed $m_{\gamma}\leq 3H/2$. After a suitable change of variables ($x\equiv -k_{1}\tau$) and more algebraic manipulations, one finds
\begin{eqnarray}
\label{111}	&& A^{\text{mix}}=P_{h}^{(0)}\cdot\frac{\pi^{2}}{32}\left(\frac{m^{2}M^{2}\Gamma}{H_{}^{2}M_{g}^{2}} \right)^{2}\cdot\frac{1}{\kappa \xi^{2}}\| \mathcal{I}_{A} \|^{2} \,,\\
\label{112}
    && B^{\text{mix}}=P_{h}^{(0)}\cdot\frac{\pi^{2}}{16}\left(\frac{m^{2}M^{2}\Gamma}{H_{}^{2}M_{g}^{2}} \right)^{2}\cdot\frac{1}{\kappa \xi^{2}}\text{Re}\left[ \mathcal{I}_{B} \right] \,,
\end{eqnarray}
where 
\begin{eqnarray}\label{Ia}
	&& \mathcal{I}_{A} \equiv \int_{x}^{+\infty}\frac{dx_{1}}{x_{1}}\,H_{\nu_{T}}^{(1)}(x_{1}) H_{\tilde{\nu}_{T}}^{(1)}(x_{1})\,,\\\label{Ib}
	&& \mathcal{I}_{B} \equiv \int_{x}^{+\infty}\frac{dx_{1}}{x_{1}}\,H_{\nu_{T}}^{(1)*}(x_{1}) H_{\tilde{\nu}_{T}}^{(1)}(x_{1})\int_{x_{1}}^{+\infty}\frac{dx_{2}}{x_{2}}\,H_{\nu_{T}}^{(1)*}(x_{2}) H_{\tilde{\nu}_{T}}^{(1)*}(x_{2}) \,.
\end{eqnarray}
\noindent The total (i.e. accounting for both polarisations) tree-level power spectrum, $P_{\rm  tot}^{}\equiv 2\left(P_{h}^{(0)}+P_{h}^{\text{mix}}\right)$, reads 
\begin{eqnarray}\label{finalps}
	P_{\rm  tot}^{}&=&2P_{h}^{(0)}\left[1+\left(\frac{m^{2}M^{2}\Gamma}{H_{}^{2}M_{g}^{2}} \right)^{2}\cdot\frac{1}{\kappa \xi^{2}}
	\frac{\pi^{2}}{32} \left(\|\mathcal{I}_{A} \|^{2}+2\text{Re}[\mathcal{I}_{B}]\right) \right]\nonumber\\&&\equiv
	2P_{h}^{(0)}\left[1+ \Theta\cdot\mathcal{C}(\nu_{T},\tilde{\nu}_{T}) \right]\,,
\end{eqnarray}
where $\mathcal{C}(\nu_{T},\tilde{\nu}_{T})\equiv \|\mathcal{I}_{A} \|^{2}+2\text{Re}[\mathcal{I}_{B}]$ and, using Eqs.~(\ref{eq:massx})-(\ref{eq:condition-I}), one has $\Theta\equiv\left(\frac{ \pi\, m_{X}^{2}}{4\sqrt{2}\,H_{}^{2}} \right)^{2}
\ll 1$ in our chosen regime. The function $\mathcal{C}$ takes on small values (see Fig.~\ref{cnunut}), hence $P_{h}^{\text{mix}}\ll P_{h}^{(0)}$.  The spectral index for $P_{\text{tot}}$ is given, to leading order, by the one obtained in Eq.~(\ref{s-index}), as we verify below:
\begin{eqnarray}\label{sin}
	n_{T}=\frac{d\ln\mathcal{P}_{\text{tot}}}{d\ln k}=\frac{d\ln\left[2\mathcal{P}_{h}^{(0)}\left(1+\Theta\cdot\mathcal{C} \right)\right]}{d\ln k}=\frac{d\ln\left[\mathcal{P}_{h}^{(0)}\right]}{d\ln k}+\frac{d\ln\left[1+\Theta\cdot\mathcal{C} \right]}{d\ln k}\,,
\end{eqnarray}
where we have introduced the reduced power spectra defined as $\mathcal{P}_{}\equiv P_{}\cdot k^{3}/(2\pi^{2})$. 
One finds 
\begin{eqnarray}\label{sinsub}
	\frac{d\ln\left[1+\Theta\cdot\mathcal{C} \right]}{d\ln k}\simeq 4\epsilon_{}\,\Theta_{}\cdot\mathcal{C}\,,
\end{eqnarray}
which is negligible compared to the contribution from $\mathcal{P}_{h}^{(0)}$, leading again to $n_{T}\simeq 3-2\nu_{T}$. Notice that the latter can also be rewritten as $n_{T}\simeq \frac{2m_{h}^{2}}{3H_{*}^{2}}-2\epsilon$, hence the tensor power spectrum is blue-(red)-tilted depending whether the mass term is larger(smaller) than the slow-roll parameter. A close inspection of our results so far reveals that as $m_{h}$ varies from smaller to larger values (and viceversa for $\epsilon$), the amplitude of the tensor-to-scalar ratio, $r\equiv \mathcal{P}_{\text{tot}}/\mathcal{P}_{\zeta}$ (with $\mathcal{P}_{\zeta}=H^{2}/(8\pi^{2}\epsilon M_{g}^{2})$), decreases whereas $n_{T}$ increases:
\begin{equation}\label{final-ps-2}
	r_{*}= 16\,\epsilon_{*}\left(\frac{\tau}{\tau_{*}} \right)^{\frac{2m_{h}^{2}}{3H^{2}_{*}}}\simeq 16\,\epsilon_{*}\,e^{-\frac{33\,m_{h}^{2}}{H_{*}^{2}}}\,,\quad\quad n_{T_{*}}\simeq \frac{2m_{h}^{2}}{3H_{*}^{2}}-2\epsilon_{*}\,,
\end{equation}
having set $N_{*}\simeq 50$. As expected, one recovers the results of standard single-field inflation, $r\simeq 16\,\epsilon$ and $n_{T}\simeq -2\epsilon$,  by setting $m_{h}=0$ in the quantities appearing in Eq.~(\ref{final-ps-2}). \\
\indent Because of the opposite effects of $m_{h}$ (and $\epsilon$) on the tensor-to-scalar ratio and on the spectral tilt, the model is able to easily accomodate fairly large values of $r$ while at the same time predicting an $n_{T}$ not too far from zero. This ensures that the tensor power spectrum is potentially detectable at CMB scales by upcoming probes such as the LiteBIRD satellite \cite{Lee2019LiteBIRD} and the new generation of ground-based experiments \cite{Hui:2018cvg,SPT-3G:2014dbx,SimonsObservatory:2018koc,CMB-S4:2016ple}, while at interferometer scales it could be visible by detectors such as the proposed  Big-Bang Observer \cite{Crowder:2005nr} or DECIGO \cite{Kawamura:2006up}.

\begin{figure}[t!]
	\centering
	\includegraphics[scale=0.26]{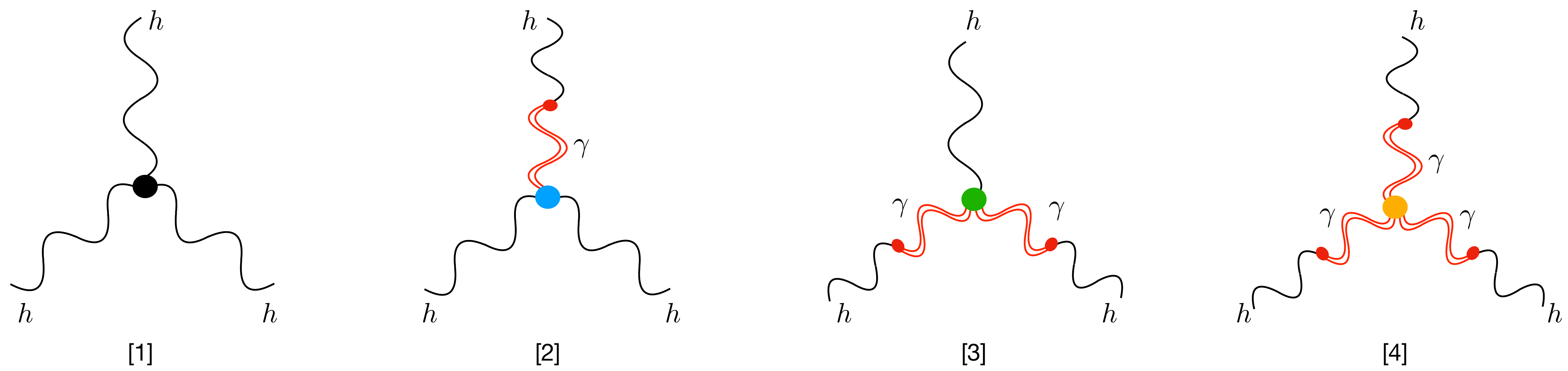}
	\caption[]{Diagrammatic representation of the tree-level contributions to the tensor bispectrum due to the interactions appearing in the third line of Eq.~(\ref{eq:action-cubic-late-time}).}
	\label{fig:BS_key}
\end{figure}

\subsection{Bispectrum}
\label{sec:b-spectra}

\noindent

\noindent The total bispectrum of $h$ at tree-level is given by the sum of the four contributions shown in Fig.~\ref{fig:BS_key}: 
\begin{equation}
	B^{\lambda_{1}\lambda_{2}\lambda_{3}}_{h}=\mathcal{A}^{\lambda_{1}\lambda_{2}\lambda_{3}}\left(B_{h}^{[1]}+B_{h}^{[2]}+B_{h}^{[3]}+B_{h}^{[4]}\right)\,,
\end{equation}
where $\mathcal{A}_{\lambda_{1}\lambda_{2}\lambda_{3}}\equiv \epsilon_{ij}^{\lambda_{1}}(\hat{k}_{1})\epsilon_{jk}^{\lambda_{2}}(\hat{k}_{2})\epsilon_{ki}^{\lambda_{3}}(\hat{k}_{3})$ can be expressed as follows 
\begin{eqnarray}
	\mathcal{A}_{\lambda_{1}\lambda_{2}\lambda_{3}}&=&-\frac{1}{64\, x_{2}^{2}\,x_{3}^{2}}\left[2x_{2}+\left(1+x_{2}^{2}-x_{3}^{2}\right)\lambda_{1}\lambda_{2} \right]\left[-2x_{3}+\left(-1+x_{2}^{2}-x_{3}^{2}\right)\lambda_{1}\lambda_{3} \right]\nonumber\\&&\quad\times\left[2x_{2}x_{3}+\left(-1+x_{2}^{2}+x_{3}^{2}\right)\lambda_{2}\lambda_{3} \right]\,,
\end{eqnarray}	
with $x_{i}\equiv k_{i}/k_{1}$, and $\lambda_{i}=\pm 1$ respectively for R,L helicities. The terms in the cubic Lagrangian responsible for each of the $B_{h}^{[i]}$ are given in the third line of Eq.~(\ref{eq:action-cubic-late-time})\footnote{The derivative interactions (first line of (\ref{eq:action-cubic-late-time})) contribute to the tree-level bispectrum with a diagram as in Fig.~\ref{fig:BS_key}-[1]. However, these interactions are not new: they are already  present  in minimal single-field slow-roll models. Further, we may not rely on the flexibility of the theory parameter space (e.g. the ratio $Q/\Gamma$) to enhance their contribution.}, which we write schematically as
\begin{equation}\label{cubic}
	\mathcal{L}^{(3)}=\alpha_{hhh}\,h^{3}+\alpha_{hh\gamma}\, h^{2}\gamma+\alpha_{h\gamma\gamma}\,h\gamma^{2}+\alpha_{\gamma\gamma\gamma}\,\gamma^{3}\,,
\end{equation}
with
\begin{eqnarray}
	&& \alpha_{hhh}  \equiv\frac{m^{2}M^{2}}{48} \left(7\,\Gamma-2\,Q \right) \,,\quad\quad\alpha_{hh\gamma} \equiv -\frac{m^{2}M^{2}}{16} \left(3\,\Gamma-2\,Q \right)   \,,\nonumber\\
	&&\alpha_{h\gamma\gamma}  \equiv-\frac{m^{2}M^{2}}{16} \left(\Gamma+2\,Q \right)    \,,\quad\quad\alpha_{\gamma\gamma\gamma}  \equiv\frac{m^{2}M^{2}}{48} \left(5\,\Gamma+2\,Q \right)    \,,
\label{Qfirst}
\end{eqnarray}
where the explicit expressions for $\Gamma$ and $Q$ are in Eqs.~(\ref{eq:GammaQdefined}). One can obtain the parametric dependence of the amplitude for the four contributions as follows:
\begin{eqnarray}\label{parametric}
	&&f_{\text{NL}}^{[1]}\propto \epsilon^{2}\,\frac{m_{h}^{2}}{H^{2}}\left(7-\frac{2Q}{\Gamma} \right) \,,\quad\quad\quad\quad\quad\quad f_{\text{NL}}^{[2]}\propto \epsilon^{2}\,\frac{m_{h}^{2}}{H^{2}}\left(3-\frac{2Q}{\Gamma} \right)\left(\frac{m_{h}/H}{\sqrt{\kappa}\xi}\right)^{2} \,,\nonumber\\
	&&f_{\text{NL}}^{[3]}\propto \epsilon^{2}\,\frac{m_{h}^{2}}{H^{2}}\left(1+\frac{2Q}{\Gamma} \right)\left(\frac{m_{h}/H}{\sqrt{\kappa}\xi}\right)^{4}  \,,\quad f_{\text{NL}}^{[4]}\propto \epsilon^{2}\,\frac{m_{h}^{2}}{H^{2}}\left(5+\frac{2Q}{\Gamma} \right)\left(\frac{m_{h}/H}{\sqrt{\kappa}\xi}\right)^{6}  \,,\nonumber\\
\end{eqnarray}
where $f_{\text{NL}}^{[i]}\propto B^{[i]}_{h}/P_{\zeta}^{2}$, $P_{\zeta}^{2}$ being the power spectrum of curvature fluctuations (see Eq.~(\ref{sq-fnl}) for the exact definition of $f_{\text{NL}}$ in the specific momentum configuration that will be relevant for our final results). The above expressions for $f_{\text{NL}}^{[2]}$, $f_{\text{NL}}^{[3]}$ and $f_{\text{NL}}^{[4]}$ have been obtained under the assumption $m_{\gamma}\leq 3H/2$. In this regime, and under the conditions on $m_{h}/H$ and $\sqrt{\kappa}\xi$ spelled out at the beginning of Sec.~\ref{sec:p-spectra} (see also Sec.~\ref{Allowed parameter space} for a full recap of these relations), one finds that $m_{h}/H\simeq \mathcal{O}(\sqrt{\kappa}\xi)$. All the quantities in Eq.~(\ref{parametric}) have therefore similar magnitudes. The complementary regime, $m_{\gamma}> 3H/2$, on the other hand, would result in oscillatory factors in the mode functions which suppress the corresponding amplitudes for $B_{h}^{[2,3,4]}$ (see e.g. \cite{Chen:2009zp}). \\

\indent In the following, we will perform a full derivation of $B_{h}^{[1]}$, i.e. the contribution from the first diagram in Fig.~\ref{fig:BS_key}. Our computations above show that it provides a very good indication of the overall level of non-Gaussianity in our model. The tree-level diagram is given by:
\begin{equation}
	\langle h_{\textbf{k}_{1}}^{\lambda_{1}}(\tau) h_{\textbf{k}_{2}}^{\lambda_{2}}(\tau)h_{\textbf{k}_{3}}^{\lambda_{3}}(\tau) \rangle^{[1]}=-i\int_{-\infty}^{\tau}d\tau'\,a(\tau')\langle\left[h_{\textbf{k}_{1}}^{\lambda_{1}}(\tau) h_{\textbf{k}_{2}}^{\lambda_{2}}(\tau)h_{\textbf{k}_{3}}^{\lambda_{3}}(\tau),H^{(3)}(\tau') \right] \rangle\,,
\end{equation}
where the brackets indicate a commutator and $H^{(3)}$ is the non-derivative interaction Hamiltonian,
\begin{equation}
	H^{(3)}(\tau')=-\alpha_{hhh}\int d^{3}\textbf{x}'\,a^{3}(\tau')\,h^{3}(\textbf{x}',\tau')\,.
\end{equation}
The latter can be expanded in Fourier space as
\begin{eqnarray}
	H^{(3)}(\tau')=&&-\alpha_{hhh}\int d^{3}\textbf{x}'\,a^{3}(\tau')\,\int\frac{d^{3}k_{1}^{'}}{(2\pi)^{3}}\int\frac{d^{3}k_{2}^{'}}{(2\pi)^{3}}\int\frac{d^{3}k_{3}^{'}}{(2\pi)^{3}}e^{-i\textbf{x}'\cdot\left(\textbf{k}_{1}+\textbf{k}_{2}+\textbf{k}_{3} \right)}\nonumber\\&&
	\times\, h_{ij,\textbf{k}_{1}}(\tau')h_{jk,\textbf{k}_{2}}(\tau')h_{ki,\textbf{k}_{3}}(\tau')\,,\nonumber
\end{eqnarray}
where
\begin{equation}
	h_{ij,\textbf{k}_{1}}(\tau')=\sum_{\lambda'}\epsilon_{ij}^{\lambda'}(\hat{k}_{1})\left[a_{\textbf{k}_{1}}^{\lambda'}h_{k_{1}}(\tau')+a_{-\textbf{k}_{1}}^{\lambda'\dagger}h_{k_{1}}^{*}(\tau') \right]\,,
\end{equation}
with commutation relations $\left[a_{\textbf{k}_{1}^{'}}^{\lambda'},a_{-\textbf{k}_{1}^{''}}^{\lambda''\dagger} \right]=(2\pi)^{3}\delta_{\lambda' \lambda''}\delta^{(3)}(\textbf{k}_{1}^{'}+\textbf{k}_{1}^{''})$. Similarly, for $\gamma$ one has 
\begin{equation}
	\gamma_{ij,\textbf{k}_{1}}(\tau')=\sum_{\lambda'}\epsilon_{ij}^{\lambda'}(\hat{k}_{1})\left[b_{\textbf{k}_{1}}^{\lambda'}\gamma_{k_{1}}(\tau')+b_{-\textbf{k}_{1}}^{\lambda'\dagger}\gamma_{k_{1}}^{*}(\tau') \right]\,,
\end{equation}
and $\left[b_{\textbf{k}_{1}^{'}}^{\lambda'},b_{-\textbf{k}_{1}^{''}}^{\lambda''\dagger} \right]=(2\pi)^{3}\delta_{\lambda' \lambda''}\delta^{(3)}(\textbf{k}_{1}^{'}+\textbf{k}_{1}^{''})$. After performing the Wick contractions and a few simplifications, the bispectrum reads
\begin{eqnarray}\label{bisgen}
	\langle h_{\textbf{k}_{1}}^{\lambda_{1}} h_{\textbf{k}_{2}}^{\lambda_{2}}h_{\textbf{k}_{3}}^{\lambda_{3}}\rangle^{[1]}&=&(2\pi)^{3}\delta^{(3)}(\textbf{k}_{1}+\textbf{k}_{2}+\textbf{k}_{3})\,12\,\mathcal{A}_{\lambda_{1}\lambda_{2}\lambda_{3}}
	\\&&\cdot\text{Im}\left[-\alpha_{hhh}\,h_{k_{1}}(\tau) h_{k_{2}}(\tau)h_{k_{3}}(\tau) \int_{-\infty}^{\tau} d\tau^{'}\,a^{4}(\tau^{'})\cdot h_{k_{1}}^{*}(\tau^{'}) h_{k_{2}}^{*}(\tau^{'})h_{k_{3}}^{*}(\tau^{'}) \right]\,,\nonumber
\end{eqnarray}
where the factor of 6 accounts for the number of permutations. 
\noindent Let us now perform a change of variables and replace the mode functions in Eq.~(\ref{bisgen}). We define $x\equiv -k_{1}\tau$, $x'\equiv -k_{1}\tau'$. One then finds 
\begin{eqnarray}
	\langle h_{\textbf{k}_{1}}^{\lambda_{1}} h_{\textbf{k}_{2}}^{\lambda_{2}}h_{\textbf{k}_{3}}^{\lambda_{3}}\rangle^{[1]}&=&(2\pi)^{3}\delta^{(3)}(\textbf{k}_{1}+\textbf{k}_{2}+\textbf{k}_{3})\,48\sqrt{2}\,\pi^{3/2}\,\mathcal{A}_{\lambda_{1}\lambda_{2}\lambda_{3}}\left(-\frac{\alpha_{hhh}H^{2}_{*}}{M_{g}^{6}} \right)\left(\frac{x}{x_{*}} \right)^{3\epsilon_{*}}x^{9/2-3\nu_{T}}\nonumber
	\\&&\cdot\text{Re}\left[\int_{x}^{+\infty}dx'\,\sqrt{x'}\,\left(\frac{x'}{x_{*}} \right)^{-\epsilon_{*}}H_{\nu_{T}}^{(1)}\left(\frac{}{}x'\right)H_{\nu_{T}}^{(1)}\left(\frac{k_{2}}{k_{1}}x'\right)H_{\nu_{T}}^{(1)}\left(\frac{k_{3}}{k_{1}}x'\right) \right]\nonumber\\&&\cdot \left(\frac{k_{2}}{k_{1}} \right)^{3-\nu_{T}}\left(\frac{k_{3}}{k_{1}} \right)^{3-\nu_{T}}\frac{k_{1}^{3}}{(k_{1}k_{2}k_{3})^{3}}\,.
\end{eqnarray}

\noindent In the regime where $m_{h}\ll 3H/2$, the bispectrum can be computed analytically and one obtains
\begin{eqnarray}\label{eqbisp0}
	\langle h_{\textbf{k}_{1}}^{\lambda_{1}} h_{\textbf{k}_{2}}^{\lambda_{2}}h_{\textbf{k}_{3}}^{\lambda_{3}}\rangle^{[1]}&=&(2\pi)^{3}\delta^{(3)}(\textbf{k}_{1}+\textbf{k}_{2}+\textbf{k}_{3})\mathcal{A}_{\lambda_{1}\lambda_{2}\lambda_{3}}\frac{H_{*}^{4}}{M_{g}^{4}} \cdot\frac{m_{h}^{2}}{H_{*}^{2}}\cdot\left(7-\frac{2Q}{\Gamma}\right)\cdot\mathcal{F}(k_{1},k_{2},k_{3})\,,\nonumber\\
\end{eqnarray}
to leading order in both $m_{h}/H$ and the slow-roll parameter, with
\begin{eqnarray}
	\mathcal{F}(k_{1},k_{2},k_{3})\equiv -\frac{2}{3 \,k_{1}^{3}k_{2}^{3}k_{3}^{3}}\left[\left(\frac{32}{3}-8\gamma  \right)\sum_{i=1}^{3}k_{i}^{3}+8\sum_{i\not= j=1}^{3}k_{i}^{2}k_{j}-8\,\prod_{i=1}^{3}k_{i}-8\,\sum_{i=1}^{3}k_{i}^{3}\cdot\ln\left(\frac{k_{t}\,x}{k_{1}}\right)\right]\,.\nonumber\\
\end{eqnarray}
The quantity $\gamma$ is here Euler’s constant and $k_{t}\equiv k_{1}+k_{2}+k_{3}$. One may verify that this bispectrum receives its largest contribution in the configuration where one of the momenta is much smaller than the other two, e.g. $k_{3}\ll k_{1}\simeq k_{2}$ (see Sec.~\ref{GW Bispectrum shape-function} for a full discussion on the shape function). In this limit, one finds
\begin{eqnarray}\label{eqbisp}
	B_{h}^{[1]}\simeq
	\frac{H_{*}^{4}}{M_{g}^{4}} \cdot\frac{m_{h}^{2}}{H_{*}^{2}}\cdot\left(7-\frac{2Q}{\Gamma}\right)\left(\frac{32\,(3\gamma+\ln(8)-7+\ln(x))}{9\,k_{1}^{3}k_{3}^{3}}  \right)\simeq -\frac{H_{*}^{4}}{M_{g}^{4}} \cdot\frac{m_{h}^{2}}{H_{*}^{2}}\left(7-\frac{2Q}{\Gamma}\right)\cdot\frac{545}{k_{1}^{3}k_{3}^{3}}\nonumber\\
\end{eqnarray}
and
\begin{eqnarray}\label{ampsfsr}
	&&\mathcal{A}_{\lambda_{1}\lambda_{2}\lambda_{3}}\simeq \begin{cases}
		\frac{1}{4} & \,\, \text{for $\,\,\lambda_{1},\,\lambda_{2},\,\lambda_{3}=\text{RRR},\,\text{LLL},\,\text{RRL},\,\text{LLR}$}\\
		\frac{1}{16}\, \frac{k_{3}^{2}}{k_{2}^{2}} & \,\, \text{for $\,\,\lambda_{1},\,\lambda_{2},\,\lambda_{3}=\text{LRR},\,\text{RLL},\,\text{RLR},\,\text{LRL}$\,.}
	\end{cases}
\end{eqnarray}
The numerical coefficient in Eq.~(\ref{eqbisp}) was obtained \cite{Endlich:2013jia} by equating $x=e^{-N_{*}}$ (i.e. choosing $k_{1}=k_{*}$), with $N_{*}\simeq 50$.\\

\noindent Upon defining the non-linearity parameter in the squeezed limit as \cite{Shiraishi:2019yux} 
\begin{equation}\label{sq-fnl}
	f_{\text{NL}}^{\lambda_{1}\lambda_{2}\lambda_{3}}\equiv \frac{B_{h}^{\lambda_{1}\lambda_{2}\lambda_{3}}(k_{1},\,k_{2},\,k_{3})|_{k_{3}\ll k_{1}\simeq k_{2}}}{(12/5)P_{\zeta}(k_{1})P_{\zeta}(k_{3})} \,,
\end{equation}
one finds for the leading contributions (i.e. those due to the RRR, RRL, LLR and LLL helicity combinations):
\begin{equation}
	f_{\text{NL}}^{[1]}\simeq - 908\,\epsilon^{2} \,\frac{m_{h}^{2}}{H^{2}}\,\left(7-\frac{2Q}{\Gamma}\right)\,.
	\label{abc}
\end{equation}
We can see from Eqs.~(\ref{abc}) and (\ref{final-ps-2}) that the amplitude of non-Gaussianity is sensitive to both the tensor-to-scalar ratio and the mass. One can safely maximise $r$ by setting it equal to $0.05$ \footnote{The current bound from CMB data, obtained under the assumption of single field slow roll inflation is $0.056$ \cite{Planck:2018jri}; this bound weakens slightly in models, such as the one under study, where the relation $n_{T}=-r/8$ no longer holds.}. The quantity $m_{h}^{2}/H_{*}^{2}$ can be chosen as large as $0.003$ \footnote{This stems from requiring $\frac{2m_{h}^{2}}{3H_{*}^{2}}\ln(x)\ll 1$, a condition necessary to prevent perturbation theory from breaking down (see e.g. \cite{Endlich:2013jia}).}. With these values for $r$ and $m_{h}^{2}/H^{2}_{*}$ in place, one obtains the following set of predictions:
\begin{equation}
\label{eq:examplefnl}
	f_{\text{NL}}^{[1]}\simeq -3\times 10^{-5}\,\left(7-\frac{2Q}{\Gamma}\right) \,,\quad \epsilon\simeq 0.0034\,,\quad n_{T}\simeq -0.0049\,,\quad H_{*}\simeq 5.8\times 10^{13}\text{GeV}\,,
\end{equation}
where the value of $H_{*}$ was obtained from the amplitude of the scalar power spectrum, $\mathcal{P}_{\zeta}\simeq 2.1\times 10^{-9}$ \cite{Planck:2018jri}. An experiment like LiteBIRD would be able to measure the tensor bispectrum with an error $\Delta F_{\text{NL}}\simeq 1$ \cite{Shiraishi:2019yux}, hence we would need a coefficient $\left\vert 7-\frac{2Q}{\Gamma}\right\vert\gtrsim 10^{5}$, for the bispectrum in our model to be observable.

\subsection{GW Bispectrum shape-function}
\label{GW Bispectrum shape-function}

\begin{figure}[h!]
	\centering
\includegraphics[scale=0.5]{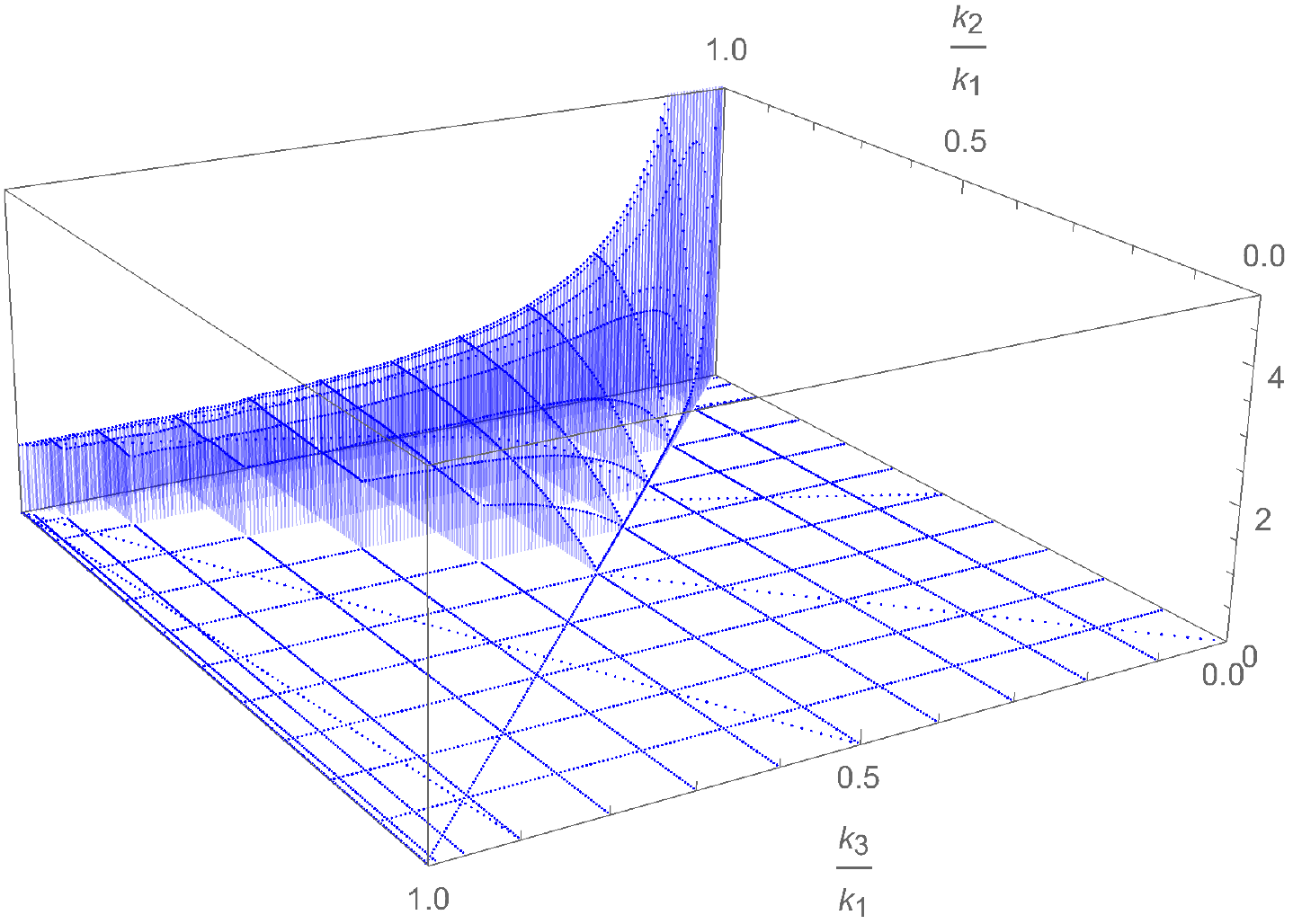}
	\caption[]{The bispectrum shape as a function of the momenta ratios $k_3/k_1$ and $k_2/k_1$. For the local template, see e.g. \cite{Celoria:2018euj} and references therein. A scalar product may be defined among different shape functions to quantify exactly their similarity \cite{Babich:2004gb}. It suffices to say here that our bispectrum is sufficiently close to the local template (i.e. a scalar product $\geq$ 0.75) to justify the comparison beyond the level of a simple inspection. 
	}
	\label{fig:shape_light}
\end{figure}

We plot in Fig.~\ref{fig:shape_light} the bispectrum as a function of momenta. As per the usual convention, the shape-function is obtained by setting one of the momenta to $1$ (given our symmetric interaction term, this can be done without loss of generality) i.e. by setting $x_i=k_i/k_1$, and multiplying the bispectrum by the factor $x_1^2 x_2^2 x_3^2$.

One can see that, for the parameter space of interest, the shape-function is peaked in correspondence of the $k_3\ll k_2$ corner of the domain (in plotting the shape function, we have made the choice $k_3\leq k_2$, without loss of generality).  This is of course very reminiscent of the so-called local template. Such finding is expected on account of two notions: (i) the contribution we are considering is that of a very light field; (ii) the specific interactions at hand are strictly non-derivative. Moving towards a massive wave-function and/or considering derivative interactions will tilt the shape towards the equilateral (or flattened, orthogonal) configuration. This interpolation between different shapes is found, for example, also in the scalar sector of quasi single-field inflation \cite{Chen:2009zp}.

\begin{figure}[h!]
	\centering
	\includegraphics[scale=0.6]{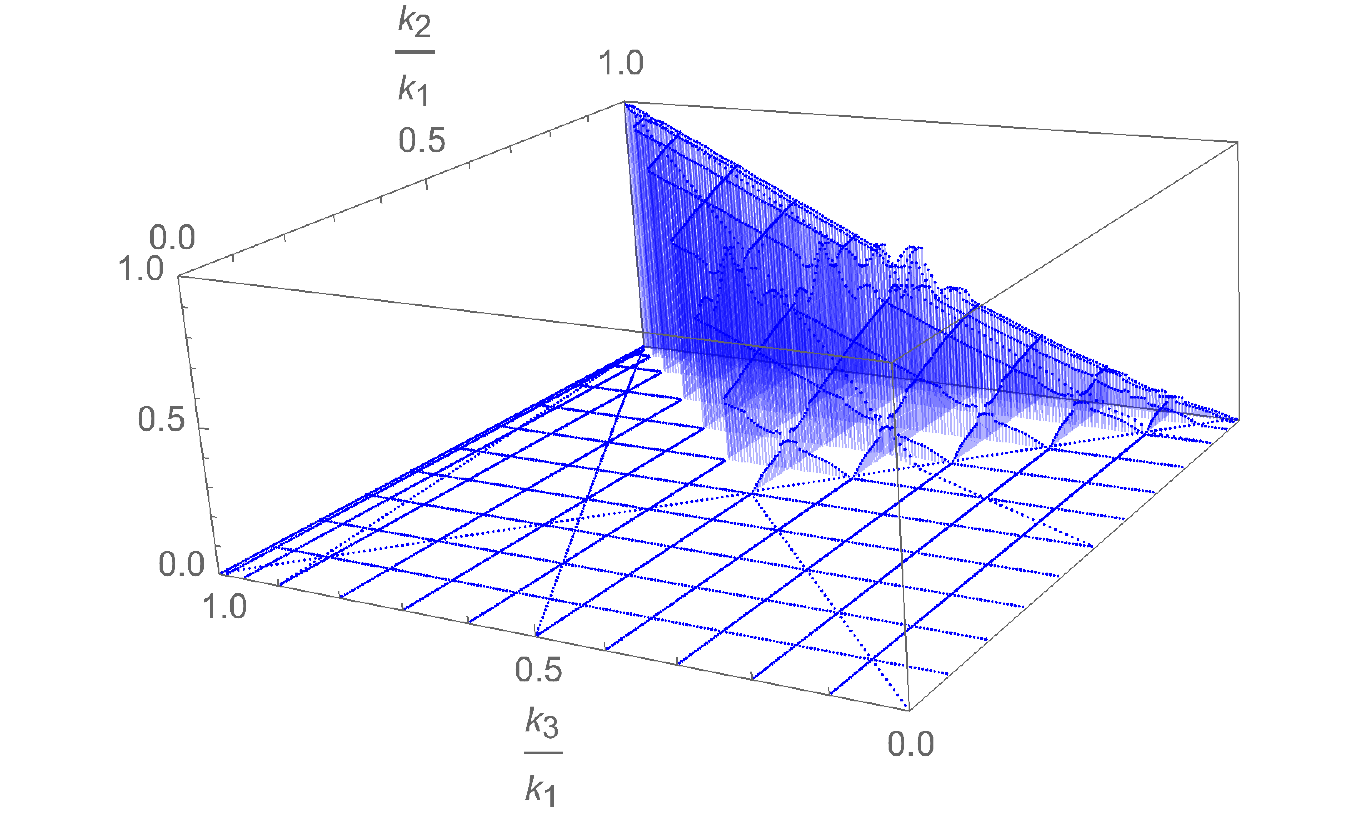}
	\caption[]{The plot of the bispectrum shape-function for $\nu_{T}$=1/2. We ought to stress again that this value of $\nu_{T}$ corresponds to an excessively massive field $h$ to satisfy all our constraints. We show this plot only as a confirmation of our line of reasoning vis-\`{a}-vis the expected shape function for sufficiently massive (i.e. $m_h\sim H$) fields. The peak of the wave-function, at the (1,1) coordinates in the momenta plane, is the same as that of the equilateral template.
	} 	\label{fig:shape_massive}
\end{figure}

The fact that the shape-function of (increasingly) derivative interactions favours equilateral, orthogonal or flat shapes (all have in common a peak that is far from the $k_3\ll k_2\sim 1$ corner is intuitively clear. Derivatives in $k$ space generate positive powers of momenta, which in turn suppress the shape-function in the squeezed limit (where the local template peaks).  For the sake of completeness\footnote{Naturally, an external field with mass of order Hubble would decay within a few e-folds. We are interested here only in confirming the expectation that a massive wave-function will ``move'' the shape towards the equilateral configuration as compared to its light or nearly massless counterpart. The explanation of such an effect is due to the $(k \tau)^{\Theta(1)}$ extra behaviour that a massive field wave-function has w.r.t. a light particle at late times. Most of the contribution from cosmological correlators is generated at horizon crossing so that one can write $(k \tau)^{\Theta(1)}\sim (k/k_{\rm fixed})^{\Theta(1)}$. These extra factors propagate all the way to the shape function and suppress the signal at small k values, i.e. in the squeezed configuration.    }, we consider in Fig.~\ref{fig:shape_massive} the case of a significantly more massive field $h$.

In reference to literature on the subject of the shape of gravitational non-Gaussianities, we should say that our model evades the conclusions in \cite{Bordin:2016ruc}
in that the results therein pertain to multiple scalar fields whilst we are dealing with multiple spin-2 particles.  The presence of the latter is crucial for the local shape to emerge. We are dealing with non-derivative interactions of interacting spin-2 fields whose massive eigenstates comprise a very light/massless field and a very massive one (with respect to the Hubble rate $H$).  The shape we find is one of the possibilities beyond the so-called partially massless case. Such additional possibility for the ``shapes of gravity'' was also listed by \cite{Goon:2018fyu} in a somewhat similar context to ours.

\section{Allowed parameter space}
\label{Allowed parameter space}
In this section, we revisit and discuss all the conditions on the parameter space stemming from the regime of validity of our approximations, theoretical consistency and observational bounds.

\subsection{Theoretical consistency}
\label{blue-p}

\paragraph{Validity of approximations and Higuchi bound}

\begin{figure}[h!]
	\centering
	\includegraphics[width=0.7\columnwidth]{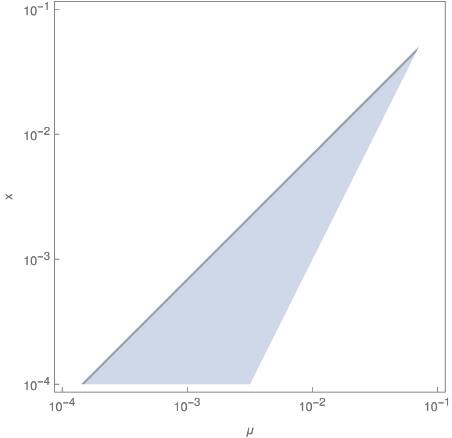}
	\caption{In the blue region, the conditions \eqref{eq:condition-AB} are satisfied. For this plot, we assumed that \textit{much less than} means \textit{less than $1/10$th of}. The thin grey line corresponds to the more restricted case where $\tilde{\nu}_{T}^2 =9/4 - \mu^2/x^2> 0$. }
	\label{fig:xmuregion}
\end{figure}

We start by defining two key dimensionless parameters:
\begin{equation}
x \equiv \sqrt{\kappa}\,\xi\,,
\qquad
\mu^2 \equiv \frac{m_h^2}{H^2} = \frac{m^2\kappa\,\Gamma}{(\kappa+1)\,H^2}\,.
\label{eq:defxmu}
\end{equation}
With these definitions, the conditions \eqref{eq:condition-I}--\eqref{eq:condition-III} can be written in the following compact form 
\begin{equation}
 \label{eq:condition-AB}
 \frac{2\,x^2}{1+x^2}  < \mu^2 \ll x \ll 1\,.\\
\end{equation}
The values of $x$ and $\mu$ allowed by this inequality are shown in Fig.\ref{fig:xmuregion}.\footnote{Note that, were we to impose real values for $\tilde{\nu}_{T}\equiv \sqrt{9/4-m_\gamma^2/H^2}$, the available region of the $(\mu,x)$ plane would be dramatically reduced, with $\mu/x \in [\sqrt{2},3/2]$. In this study however, we allow also for an imaginary $\tilde{\nu}_{T}$ for simplicity. In this last case, the correlators are unaffected by the more massive tensor degree of freedom as it decays away, hence the first diagram in Fig.~\ref{fig:BS_key} remains the dominant contribution to the bispectrum.
}

\paragraph{Conservative cutoff of the EFT description}
Ours is an interacting spin-2 model that is often dubbed a $\Lambda_3$ theory because its na\"ive cutoff is of the type $(m^2 M_f)^{1/3}$ or $(m^2 M_g)^{1/3}$, whichever quantity is the smallest. Our constraints here are conservative because, depending on the background, the cutoff is typically revisited towards higher values \cite{deRham:2010kj}.

 Assuming $\kappa<1$, we require that the energy scale at which we are operating during inflation, i.e. $H$, does not exceed the cutoff, i.e.
\footnote{The assumption of $\kappa<1$ provides the most conservative conditions. If $\kappa >1$, the lowest scale becomes $(m^2M_g)^{1/3}$ and the condition to remain within the consistent EFT regime becomes weaker.}
\begin{equation}
H  < (m^2 M_f)^{1/3}\,.
\end{equation}
Since we will later fix $H$ in order to (almost) saturate the bound on the tensor-to-scalar ratio $r$ \cite{Planck:2018jri}, this condition can be interpreted as a bound on $m$
\begin{equation}
m^2 > \frac{H^3}{M_g\,\sqrt{\kappa}}\,.
\label{eq:constraint-EFT}
\end{equation}

\paragraph{Existence of the low energy regime}
The cosmological solutions for two interacting spin--2 fields satisfy $\tilde{c}=1$ and $\dot{\xi}=0$ in the regime where
\begin{equation} 
\rho \ll m^2 M^2\,.
\end{equation}
In order to reach this regime already during inflation, we ought to impose another lower bound on $m$, namely
\begin{equation}
m^2 \gg \frac{3\,H^2\,(1+\kappa)\,(1+x^2)}{\kappa}\; .
\label{eq:constraint-LATE}
\end{equation}
This condition is stronger than the condition on the validity of EFT \eqref{eq:constraint-EFT} so long as $H < M_p$ and $x\ll 1$, which, as we shall see, are both desirable and well justified in our scenario. By using the definition of $\mu$ \eqref{eq:defxmu}, we can reformulate Eq.\eqref{eq:constraint-LATE} to give an upper bound on a combination of the interaction parameters $\beta_n$ as 
\begin{equation}
\Gamma\ll \frac{\mu^2}{3}\,,
\label{eq:Gammaupper}
\end{equation}
where we approximated $1+x^2\simeq 1$ in accordance with \eqref{eq:finalc-xm}, a regime we are going to discuss soon. 

\paragraph{Perturbative Stability}
In order to ensure the absence of scalar and vector instabilities, as detailed in \eqref{eq:stability-late} and \eqref{eq:stability-early}, one ought to impose
\begin{equation}
 \Gamma>0\,,\qquad
 \beta_1 >0\,.
 \label{eq:condition-STAB}
\end{equation}
We remind the reader that in order to prevent an effective cosmological constant arising from the two-metric interaction, we fix the mass parameters so that
\begin{equation}
\rho_{m,g}=\rho_{m,f}=0\,.
 \label{eq:condition-LATE2}
\end{equation}

\subsection{Consistency with observations}

\paragraph{Gravitational coupling constants} The gravitational constant for cosmology \eqref{eq:Meff} differs from the gravitational constant for Newton's law, which is determined solely by the coupling constant $M_g$ (see e.g.\cite{Fasiello:2015csa}).  Since the value of $G_N$ is determined observationally with $\mathcal{O}(10^{-5})$ relative uncertainty \cite{Zyla:2020zbs}, we fix
\begin{equation}
M_g = M_p\,.
\end{equation}
In cosmology however, the effective coupling constant could affect the freeze-out time of weak interactions and introduce a modification to the primordial Helium abundance \cite{Carroll:2004ai}
\begin{equation}
\Delta Y_p = 0.08 \,\left(\frac{G_{C}}{G_N} - 1\right)\,.
\end{equation}
Recent observations of $Y_p$ \cite{Izotov:2014fga,Aver:2015iza,Valerdi:2019beb} have an uncertainty $|\Delta Y_p| < 0.01 \, (99.7\%\,\, {\rm C.L.})$\footnote{We use the uncertainty suggested by the PDG in Ref.\cite{ParticleDataGroup:2020ssz}}, which constrains the deviation of the Planck masses at
\begin{equation}
\left\vert\frac{M_p^2}{M_{eff}^2} - 1\right\vert  = \frac{x^2}{1+x^2}<\frac18 \,,
\end{equation}
which is consistent with the results of the recent study \cite{Hogas:2021saw}. The region where this condition is satisfied is
\begin{equation}
 x < \frac{1}{\sqrt{7}}\,,
 \label{eq:condition-BBN}
\end{equation}
which is already implemented via the last inequality of \eqref{eq:condition-AB}.

\paragraph{Solar system tests}
The scalar polarisation of the graviton can affect local tests of gravity. It is possible to evade such tests whenever the Vainshtein screening mechanism is active. Screening is very effective in the following regime \cite{DeFelice:2013nba,DeFelice:2014nja,Kenna-Allison:2018izo}
\begin{equation}
\left\vert\frac{d\,\log \Gamma}{d\,\log \xi}\right\vert \gg 1\,,
\end{equation}
a condition that simply gives
\begin{equation}
\left\vert \frac{Q}{\Gamma} - 1 \right\vert  \gg 1\,.
\end{equation}
In the regimes where the tensor bispectrum \eqref{abc} is observable, i.e. $f_{\text{NL}}^{[1]}\sim \mathcal{O}(1)$, the ratio $|Q/\Gamma|$ is already very large so that the local tests of gravity are impervious to the scalar graviton without the need for any extra assumption.

\paragraph{GW observations}
LIGO observations of gravitational waves from binary mergers have set an upper bound on the mass of the tensor mode in theories with interacting spin-2 particles. If we were to assume that the full spin-2 field content active during inflation survives all the way to today, such constraint would affect our parameter space via \cite{LIGOScientific:2016lio}
\begin{equation}\label{ww}
m_h < 10^{-22} {\rm eV} \sim 10^{-49} M_{p}\,.
\end{equation}
Using the relation $H\simeq 10^{15}\,\sqrt{\epsilon}$, obtained from the amplitude of the scalar power spectrum, and rewriting (\ref{ww}) as a constraint on $\mu$, one would have
\begin{equation}
\mu \sqrt{\epsilon} < {10^{-46}}\,.
\label{eq:constraint-GW}
\end{equation}
Such bound would be very demanding on our parameter space, either on $\mu$ (i.e. the mass of the graviton) or on $\epsilon$. In the latter case we would have a very flat inflationary potential, to the detriment of any observational prospect for the GW spectrum. On the other hand, applying the LIGO bound would necessarily mean presupposing that the additional ``$f$'' spin-2 sector has survived until the present time without interacting with any other field. It is perhaps more realistic to assume instead that this massive spin-2 particle does interact with other field content and may decay some time after inflation.

One potential realisation for such a scenario 
is to allow the spin-2 interaction parameter $m$ to vary with the dynamics of a scalar field. By dialling down the interaction in the post-inflationary universe, it is possible to have tensors that are massive at early times and are hardly affected by the LIGO bound at the present time. Such a mechanism was introduced in Ref.~\cite{Hinterbichler:2013dv} in the context of massive spin-2 theories with varying mass \cite{Huang:2012pe}. With these considerations in mind, in the remainder of the paper we will disregard the constraints on the tensor mass from late-time gravitational wave observations.

\section{Discussion: summary of the constraints and a concrete example}
\label{discussion}

We summarise the constraints obeyed by the parameters of the model under the additional requirement that the theory generate observable GW non-Gaussianity. The conditions arising from the validity of approximations \eqref{eq:condition-AB} and BBN \eqref{eq:condition-BBN},
are simultaneously satisfied for
\begin{equation}
2\,x^2 < \mu^2 \ll x\ll 1/\sqrt{7}\,.
\label{eq:finalc-xm}
\end{equation}

Before moving on to the conditions on the $\beta_n$ parameters, we first make use of specific relations to reduce the number of free parameters. Since the absence of an effective cosmological constant at late times \eqref{eq:condition-LATE2} imposes $\rho_{m,g}=\rho_{m,f}=0$, we can fix two of the parameters using the definitions in Eq.\eqref{eq:GammaQdefined}
\begin{align}
\beta_2\xi^2 =& -\frac16\,\left(3\,\beta_0+8\,\beta_1\xi-\beta_4\xi^4\right)\,,\nonumber\\
\beta_3\xi^3=& \frac12\,\left(\beta_0+2\,\beta_1\xi-\beta_4\xi^4\right)\,.
\label{eq:eqb2b3}
\end{align}
With this specification, $\beta_0$ and $\beta_4$ can be written as:
\begin{align}
\beta_0 =& Q-2\,\Gamma-\beta_1\xi\,,\nonumber\\
\beta_4\xi^4=& -3\,Q-\beta_1\xi\,,
\label{eq:eqb1b4}
\end{align}
thus reducing the freedom in the $\beta_n$ parameters to the one in $\beta_1$, $Q$ and $\Gamma$. The constraints on these quantities can be summarised by combining the validity of the late-time attractor condition \eqref{eq:Gammaupper}, the perturbative stability conditions \eqref{eq:condition-STAB}, and the requirement for an observable tensor bispectrum \eqref{abc}:
\begin{equation}
0 < \Gamma \ll \frac{\mu^2}{3}\,,\qquad \beta_1 >0\,,\qquad \left\vert\frac{Q}{\Gamma}\right\vert \gtrsim \frac{1}{r^2\,\mu^2}\,.
\label{eq:finalc-betas}
\end{equation}

Let us provide a concrete example of an interesting parameter space region in our model. We consider the values discussed in \eqref{eq:examplefnl} with $r=0.05$ and the mass at $\mu^2=3\times10^{-3}$. Next, in compliance with \eqref{eq:finalc-xm}, we fix $x=10^{-2}$.
True to Eq.\eqref{eq:finalc-betas}, we consider the value $\Gamma = 10^{-5}$. For the other two constrained parameters, we take $|Q/\Gamma| = 10^{5}$ and $\beta_1=1$. In order to arrive at specific numbers, we set $\kappa=1/9$, which implies $\xi=0.03$. The hierarchy between $Q$ and $\Gamma$ is then obtained by judiciously choosing the $\beta_n$ parameters. There is one subtlety associated to the choice of any value for $\xi$\footnote{Although Eqs.\eqref{eq:eqb2b3} and \eqref{eq:eqb1b4} always produce parameters that do satisfy the late-time equation \eqref{eq:eqXbetas-late}, one ought to verify that the solution used for the late-time value of $\xi$ is the first root to be reached. This is because the early time evolution starts from $\xi\sim 0$ and therefore only the smallest positive root of Eq.\eqref{eq:eqXbetas-late} will be dynamically accessible. The reason is that when Eq.\eqref{eq:eqXbetas-late} is satisfied, $\xi$ stops evolving. We find that, upon requiring  $Q<0$, a set of parameters consistent with Eq.\eqref{eq:condition-LATE2} can be readily obtained.}. With that taken care of, we may obtain $Q/\Gamma = -10^{5}$ by for example employing the following $\beta_n$ values: $\beta_0 = -100302\times 10^{-4}\;;\beta_1 \xi=3\times 10^{-2}\;;
\beta_2 \xi^2= 99701 \times 10^{-4}\;;
\beta_3 \xi^3= -199701\times 10^{-4}\;;
\beta_4 \xi^4 = 2997\times 10^{-2}\,$, a choice for which  the smallest positive root of Eq.\eqref{eq:condition-LATE2} is indeed precisely $\xi=0.03$.

It is important to remark here that
the specific bispectrum contribution we calculate does break so-called consistency relations (CRs) and is therefore directly physical.  In order to see that this is indeed the case, we can rely on a parametric proof. It suffices to show that the tensor power spectrum may not contain a certain ``free'' (or -adjustable- within a interval of values set by e.g. observational or unitarity constraints) parameter that appears instead in the squeezed GW bispectrum.
Now, in our case it is the parameter $Q$, see e.g. Eq.~(\ref{Qfirst}), (as opposed to the parameter $\Gamma$, which also regulates the bispectrum but does not share this property) that plays this role: it appears in the bispectrum but, upon analysing the quadratic tensor Lagrangian (as well its quadratic corrections via the in-in formalism, see Eqs.~\ref{111} and \ref{112}), it becomes clear it may not appear in the power spectrum.

The parameter space region corresponding to the set of values we just analysed will deliver GW non-Gaussianity that is within reach for both the LiteBIRD satellite mission and CMB-S4 experiments \cite{Shiraishi:2019yux}. Next generation CMB probes provide an excellent handle on the early universe particle content. The same is in general true for probes of gravity at much smaller scales, including laser interferometer. For the purposes of testing the model at hand though, it is large scales that will deliver the most stringent constraints.

\section{Conclusions}
\label{sec:conclusions}

Current and next generation cosmological probes will put our ideas on the early (and late) universe to the test. In this paper we have focused on an inflationary model with a rich field content, going beyond the minimal single-field slow-roll scenario. It is well-known that we may test for a rich inflationary content across all scales. At large scales, CMB experiments as well as astronomical surveys of the galaxy distribution promise to zoom in on inflationary (self)interactions by setting strong constraints on non-Gaussianity parameters.

 The improvements with respect to existing bounds will be especially remarkable for gravitational wave non-Gaussianities, with a projected $\sigma_{f^{\rm tens}_{\rm NL}}\sim 1$ from both LiteBIRD and CMB-S4 experiments.
Large scale structure constraints (e.g. from Euclid, SKA, LSST) will also reach $\sigma_{f^{\rm }_{\rm NL}}\sim 1$ for the scalar sector. Remarkably, testing gravitational waves at  intermediate and small frequencies is also guaranteed to teach us a lot about the early universe. There exist several cosmological sources of primordial gravitational waves, from those produced during pre-heating dynamics, to those associated to first order phase transitions, to GWs due to energy loss from cosmic strings. However, perhaps no  cosmological mechanism may claim GWs as a universal prediction more than inflation.

Given that the minimal single-field slow-roll paradigm is typically associated with a slightly red-tilted GW spectrum, a mere GW detection\footnote{ That is, the detection of a GW event whose source may be clearly identified as being of cosmological rather than astrophysical origin.} in the intermediate (i.e. within reach of pulsar timing arrays) and small frequency range (i.e.  in the LISA, LIGO, Einstein Telescope, Taiji, Cosmic Explorer and BBO band) is generally seen as a smoking gun for multi-field dynamics.

The model we studied in this work however, does not exhibit a GW spectrum that is detectable at small scales. To identify and test its most salient features, we relied on its large scale signatures. The field content consists of minimal ingredients (massless spin-2 + scalar) with the addition of an ``extra'' spin-2 field. Spinning fields are particularly interesting in view, for example, of their non-analytic momentum scaling and extra angular behaviour associated to their (squeezed) non-Gaussian signal \cite{Arkani-Hamed:2015bza}. Starting with spin-2 particles, it is also possible to source GW already at linear level.  

The cosmological imprints of inflationary fields with multiple spin-2 content have been tackled in the literature from several points of view. It is possible, for example \cite{Bordin:2018pca,Iacconi:2019vgc,Iacconi:2020yxn,Malhotra:2020ket}, to rely on an EFT description that assumes a well-behaved FRW background and starts from a Lagrangian from fluctuations atop the classical solution.   A more conservative approach, more in the spirit of our present work, consists in employing the only available fully non-linear theory of interacting spin-2 fields in the inflationary context (see e.g. \cite{Biagetti:2017viz,Dimastrogiovanni:2018uqy}).  

Our work in this manuscript has gone beyond the results in \cite{Biagetti:2017viz,Dimastrogiovanni:2018uqy} in manifold ways. One of the key differences is our use of the full FRW solutions (\textit{vis-\`{a}-vis} the de Sitter approximation found in previous works) for the Lagrangian, which revealed a much wider parameter space, sufficiently flexible to grant detectable levels of GW non-Gaussianity. Our thorough study of constraints stemming from theoretical consistency of the FRW solution is also a novel addition in the inflationary context.
 
It remains to be seen whether, upon considering non-minimal coupling to the inflaton, the same field content may produce a GW signal that is testable also at small scales. This is an interesting direction we plan to pursue in the future, full in the knowledge that direct consistent coupling to ``matter'' is possible only under rather stringent  conditions \cite{deRham:2014naa,deRham:2014fha} in the context of fully non-linear models.

\acknowledgments

MF would like to acknowledge support from the ``Atracci\'{o}n de Talento'' grant 2019-T1/TIC15784. AEG is supported by a Dennis Sciama Fellowship at the University of Portsmouth.

\appendix
\section{Perturbed action up to cubic order}
\label{app:allterms}

Using FLRW ansatze for both background metrics \eqref{eq:background-metric}, we introduce tensor perturbations via:
\begin{align}
ds_g^2 &= -N^2dt^2  + a^2 \left(\delta_{ij}+2\,h_{ij}\right)dx^idx^j\,,\nonumber\\
ds_f^2 &= \xi^2 \left[-N^2\tilde{c}^2 dt^2  + a^2 \left(\delta_{ij}+2\,\gamma_{ij}\right)dx^idx^j\right]\,,
\end{align}
where $h_{ij}$ and $\gamma_{ij}$ are transverse and traceless with respect to the 3-d metric $\delta_{ij}$.

\subsection{Kinetic terms}
We calculate the Einstein-Hilbert terms for the two metrics as
\begin{align}
\sqrt{-g}R[g] =& a^3\left[\mathcal{L}_{g{\rm -kin}}^{(0)}+\mathcal{L}_{g{\rm -kin}}^{(2)}+\mathcal{L}_{g{\rm -kin}}^{(3)}+\dots\right]\,,
\nonumber\\
\sqrt{-f}R[f] =& a^3\xi^4\tilde{c}\left[\mathcal{L}_{f{\rm -kin}}^{(0)}+\mathcal{L}_{f{\rm -kin}}^{(2)}+\mathcal{L}_{f{\rm -kin}}^{(3)}+\dots\right]\,,
\end{align}
where
\begin{align}
\mathcal{L}_{g{\rm -kin}}^{(0)} \equiv &
-\frac{6\,\dot{a}^2}{a^2N}\,,
\nonumber\\
%%%%%%%%%%%
\mathcal{L}_{g{\rm -kin}}^{(2)} \equiv &
\frac14\,\dot{h}^{}{}_{ij} \dot{h}^{ij} 
- \left(\dot{H}+\frac32\,H^2\right) h^{}{}_{ij} h^{ij}
- \frac{1}{4\,a^2}\,\partial_{k}h^{ij} \partial^{k}h^{}{}_{ij}
\,,
\nonumber\\
%%%%%%%%%%%
\mathcal{L}_{g{\rm -kin}}^{(3)} \equiv &
-\frac{1}{2}\,h^{ij} \dot{h}^{}{}_{i}{}^{k} \dot{h}^{}{}_{jk} 
+ \left(\frac23\,\dot{H}+H^2\right)\, h^{}{}_{i}{}^{k} h^{ij} h^{}{}_{jk}
+ \frac{1}{4\,a^2} \,h^{ij}\left[
 \partial_{i}h^{kl} \partial_{j}h^{}{}_{kl} 
- 2 \, \partial_{j}h^{}{}_{kl} \partial^{l}h^{}{}_{i}{}^{k}
+ 2 \, \partial_{l}h^{}{}_{jk} \, \partial^{l}h^{}{}_{i}{}^{k} \right]
\,,
\nonumber\\
\mathcal{L}_{f{\rm -kin}}^{(0)} \equiv &
-\frac{6\,(\xi\,\dot{a}+a\,\dot{\xi})^2}{a^2\tilde{c}^2N\,\xi^4}\,,
\nonumber\\
\mathcal{L}_{f{\rm -kin}}^{(2)} \equiv &\frac{1}{4\,\tilde{c}^2\xi^2}
\dot{\gamma}^{}{}_{ij} \dot{\gamma}^{ij} 
- \left(\frac{\dot{H}_f}{\tilde{c}\,\xi}+\frac32\,H_f^2\right) \gamma^{}{}_{ij} \gamma^{ij}
- \frac{1}{4\,a^2\xi^2}\,\partial_{k}\gamma^{ij} \partial^{k}\gamma^{}{}_{ij}
\,,
\nonumber\\
\mathcal{L}_{f{\rm -kin}}^{(3)} \equiv &
-\frac{1}{2\,\tilde{c}^2\xi^2} \gamma^{ij} \dot{\gamma}^{}{}_{i}{}^{k} \dot{\gamma}^{}{}_{jk} 
+ \left(\frac23\,\frac{\dot{H}_f}{\tilde{c}\,\xi} +H_f^2\right) \gamma^{}{}_{i}{}^{k} \gamma^{ij} \gamma^{}{}_{jk}
\nonumber\\
&
+\frac{1}{4\,a^2\xi^2}\,\gamma^{ij}\left[
 \partial_{i}\gamma^{kl} \partial_{j}\gamma^{}{}_{kl} 
- 2 \, \partial_{j}\gamma^{}{}_{kl} \partial^{l}\gamma^{}{}_{i}{}^{k}
+ 2 \, \partial_{l}\gamma^{}{}_{jk} \, \partial^{l}\gamma^{}{}_{i}{}^{k} 
\right]
\,,
\label{eq:actionexpand-kinetic}
\end{align}
where the above relations were obtained by adding appropriate boundary terms and we used the Cayley-Hamilton theorem for a traceless $3\times 3$ tensor, 
\begin{equation}
(h^3)_{ij} = \frac{[h^2]}{2}\,h_{ij}+\frac{[h^3]}{3}\,\delta_{ij}\,,
\end{equation}
with square brackets denoting the trace operation.
A similar expression applies to $\gamma_{ij}$.
We also defined the Hubble rates for the two metrics as
\begin{equation}
H \equiv \frac{\dot{a}}{a\,N}\,,
\qquad
H_f\equiv \frac{1}{\tilde{c}\,\xi}\left(\frac{\dot{a}}{a\,N}+\frac{\dot{\xi}}{\xi\,N}\right)\,.
\end{equation}
We note that the lapse function $N$ is kept only for the background action, as in the main text we calculate the equations of motion in the mini-superspace approximation. For the rest of the terms, we fix the time coordinate by choosing $N=1$.

Although we are after the $f$--$g$ interaction terms, the expansion-induced self-interaction terms from the Einstein Hilbert terms may become relevant since $h_{ij}$ and $\gamma_{ij}$ are not mass eigenstates. Moreover, we expect (at least) part of the two-metric interaction to simplify on shell when combined with the kinetic terms.

\subsection{2--metric interaction terms}

We now calculate the contribution from the $g$--$f$ interaction. We define,
\begin{equation}
\sqrt{-g}\sum_{n=0}^4 \beta_n\,e_n(\sqrt{g^-1f}) = 
a^3\left[\mathcal{L}_{{\rm int}}^{(0)}+\mathcal{L}_{{\rm int}}^{(2)}+\mathcal{L}_{{\rm int}}^{(3)}+\dots\right]\,,
\end{equation}
where
\begin{align}
\mathcal{L}_{{\rm int}}^{(0)} =&
N\,\left(\rho_{m,g}+\tilde{c}\,\xi^4\,\rho_{m,f}\right)\,,
\nonumber\\
%%%%%%%%%%%
\mathcal{L}_{{\rm int}}^{(2)} =&
-\frac{(\tilde{c}-1)}{4}\,\Gamma \,\left([h^2]-[\gamma^2]\right)
-\frac14\,\left(\rho_{m,g}\,[h^2]+\tilde{c}\,\xi^4\rho_{m,f} [\gamma^2]\right)
+\frac{[\Gamma+(\tilde{c}-1)Q]}{8}[(h-\gamma)^2]\,,
\nonumber\\
%%%%%%%%%%%
\mathcal{L}_{{\rm int}}^{(3)} =&
\frac{\beta_1(\tilde{c}-1)\,\xi}{24}\,[(h-\gamma)^3]
-\frac{\Gamma}{8}\,\left(
\frac16\,[(h-\gamma)^3]+ [(h-\gamma)^2.(h+\gamma)]-(\tilde{c}-1)\,[(h-\gamma).(h+\gamma)^2]
\right)
\nonumber\\
&
+\frac{1}{6}\left(\rho_{m,g} [h^3]+\tilde{c}\,\xi^4\rho_{m,f}[\gamma^3]\right)
+\frac{Q}{8}\,\left(
\frac{3\,\tilde{c}-1}{6}\,[(h-\gamma)^3]-(\tilde{c}-1)[(h-\gamma)^2.(h+\gamma)]
\right)\,.
\label{eq:actionexpand-interaction}
\end{align}
In the above, we used the definitions \eqref{eq:GammaQdefined} to simplify certain combinations of the $\beta_n$ parameters. As in the previous section, we keep the lapse $N$ only for the background contribution and set it to unity for higher order terms.

\subsection{Matter sector}
For the matter sector, we use a single scalar field with the general action
\begin{equation}
 S = \int d^4x \sqrt{-g} \,P(X,\phi)\,,
\end{equation}
where $X\equiv -\frac12\partial_\mu\phi \, \partial^\mu\phi$, which reduces to $X = \frac{\dot{\phi}^2}{2\,N^2}$ when only tensor perturbations are included. In analogy with a perfect fluid,  $P(X,\phi)$ corresponds to the pressure while the energy density and sound speed are defined as:
\begin{equation}
\rho = 2\,X\,P_{,X} -P\,,
\qquad 
c_s^2 = \frac{P_{,X}}{2\,X \,P_{,XX}+P_{,X}}
\,,
\label{eq:density-soundspeed}
\end{equation}
with
\begin{equation}
 P_{,X} \equiv \frac{\partial\,P(X,\phi)}{\partial X}\,,\qquad
P_{,XX} \equiv \frac{\partial^2\,P(X,\phi)}{\partial X^2}\,,\qquad
  P_{,\phi}\equiv \frac{\partial\,P(X,\phi)}{\partial \phi}\,.
\end{equation}
Below are some relations that we use to derive the background equations in 
Sec.~\ref{sec:background}:
\begin{align}
\frac{\partial P}{\partial \dot\phi} =& \frac{P+\rho}{\dot\phi}\,,
\nonumber\\
\frac{\partial P}{\partial N} =&  -\frac{P+\rho}{N}\,,
\nonumber\\
\frac{1}{N}\,\frac{d}{dt}\left(\frac{\dot{\phi}}{N}\right) = &
\frac{c_s^2 \,\dot\phi}{N^2(P+\rho)}\left(\dot\rho-\dot\phi\,\rho\right)\,,
\nonumber\\
\dot{P}=&\dot\phi\,P_{,\phi}+c_s^2\left(\dot\rho-\dot\phi\,\rho\right)\,.
\label{eq:useful-matter}
\end{align}

We thus have
\begin{equation}
\sqrt{-g}\,P(X,\phi) = a^3N\left(1-\frac{1}{4}\,[h^2]+\frac{1}{6}\,[h^3]
+\dots\right)P\,.
\label{eq:actionexpand-matter}
\end{equation}

\subsection{Total action}
We now combine all the contributions from the kinetic 
\eqref{eq:actionexpand-kinetic}, interaction \eqref{eq:actionexpand-interaction} and \eqref{eq:density-soundspeed} and calculate the total action up to cubic order in tensor perturbations
\begin{equation}
S= \int d^4x \,a^3 \left(\mathcal{L}_{\rm tot}^{(0)}+
\mathcal{L}_{\rm tot}^{(2)}+\mathcal{L}_{\rm tot}^{(3)}+\dots
\right)\,,
\end{equation}
where the explicit expression for the background is presented in Eq.~\eqref{eq:minisuperspace} and we have
\begin{align}
{\mathcal L} ^{(2)}_{\rm tot} =& \frac{M_g^2}{8}\,\left(\dot{h}_{ij}\dot{h}^{ij}-\frac{1}{a^2}\partial_kh^{ij}\partial^k h_{ij}\right)
+\frac{M_f^2\xi^2\tilde{c}}{8}\,\left(\frac{1}{\tilde{c}^2}\dot{\gamma}_{ij}\dot{\gamma}^{ij}-\frac{1}{a^2}\partial_k\gamma^{ij}\partial^k \gamma_{ij}\right)
\nonumber\\
&
-\frac{m^2M^2}{8} \,\left[\Gamma +(\tilde{c}-1)\,Q\right]\,(h_{ij}-\gamma_{ij})(h^{ij}-\gamma^{ij})\,.
\label{eq:action-quadratic}
\\
{\mathcal L} ^{(3)}_{\rm tot} = &
-\frac{M_g^2}{4} h_{ij}\left[
\dot{h}^{ik}\dot{h}_k^{\;\;j}-\frac{1}{2\,a^2}
\Big(
 \partial^{i}h^{kl} \partial^{j}h^{}{}_{kl} 
- 2 \, \partial^{i}h_{kl} \partial^{l}h^{jk}
+ 2 \, \partial_{l}h^{i}_{\;\;k} \, \partial^{l}h^{jk} 
\Big)
\right]
\nonumber\\
&-\frac{M_f^2\xi^2\,\tilde{c}}{4}\,\gamma_{ij}\left[
\frac{1}{\tilde{c}^2}\,\dot{\gamma}^{ik}\dot{\gamma}_k^{\;\;j}-\frac{1}{2\,a^2}
\Big(
 \partial^{i}\gamma^{kl} \partial^{j}\gamma^{}{}_{kl} 
- 2 \, \partial^{i}\gamma_{kl} \partial^{l}\gamma^{jk}
+ 2 \, \partial_{l}\gamma^{i}_{\;\;k} \, \partial^{l}\gamma^{jk} 
\Big)
\right]
\nonumber\\
&
+\frac{m^2M^2}{48}\left(\Gamma-(\tilde{c}+1)\,Q+2\,\xi^2(\tilde{c}-1)\beta_2\right)[(h-\gamma)^3]
\nonumber\\
&
+\frac{m^2M^2}{8}\left(\Gamma+(\tilde{c}-1)\,Q\right)\,[(h-\gamma)^2(h+\gamma)]\,.
\label{eq:action-cubic}
\end{align}

\bibliographystyle{JHEP}
\bibliography{spin2inflation}
\end{document}